\documentclass{ieeeaccess}
\usepackage{cite}
\usepackage{amsmath,amssymb,amsfonts}
\usepackage{algorithmic}
\usepackage{graphicx}
\usepackage{textcomp}
\usepackage{setspace} 
\usepackage{graphicx} 
\usepackage{array}
\newcolumntype{L}[1]{>{\raggedright\let\newline\\\arraybackslash\hspace{0pt}}m{#1}}
\newcolumntype{C}[1]{>{\centering\let\newline\\\arraybackslash\hspace{0pt}}m{#1}}
\newcolumntype{R}[1]{>{\raggedleft\let\newline\\\arraybackslash\hspace{0pt}}m{#1}}
\usepackage{booktabs}
\usepackage{amssymb}
\usepackage{pifont}
\usepackage{rotating}
\usepackage{lipsum}
\usepackage{setspace}
\usepackage{authblk}
\usepackage{multirow}

\usepackage[hyphens]{url}
\usepackage[hidelinks]{hyperref}
\hypersetup{breaklinks=true}
\usepackage{rotating}
\usepackage{pdflscape}

\def\BibTeX{{\rm B\kern-.05em{\sc i\kern-.025em b}\kern-.08em
    T\kern-.1667em\lower.7ex\hbox{E}\kern-.125emX}}
\begin{document}
\history{Date of publication xxxx 00, 0000, date of current version xxxx 00, 0000.}
\doi{XX.XXXX/XX.2018.DOI}

\title{Fog Computing: Survey of Trends, Architectures, Requirements, and Research Directions}
\author{\uppercase{Ranesh Kumar Naha}\authorrefmark{1}, 
{\uppercase{Saurabh Garg}\authorrefmark{1} \IEEEmembership{Member, IEEE}, \uppercase{Dimitrios Georgekopolous}\authorrefmark{2} \IEEEmembership{Member, IEEE}, \uppercase{Prem Prakash Jayaraman} \authorrefmark{2} \IEEEmembership{Member, IEEE},\uppercase{Longxiang Gao} \authorrefmark{3} \IEEEmembership{Senior Member, IEEE}, Yong Xiang\authorrefmark{3} \IEEEmembership{Senior Member, IEEE}, \uppercase{and Rajiv Ranjan}}\authorrefmark{4} \IEEEmembership{Senior Member, IEEE}}

\address[1]{School of Engineering and ICT, University of Tasmania, Sandy Bay, Tasmania, Australia.}
\address[2]{School of Software and Electrical Engineering, Swinburne University of Technology, Melbourne, Australia.}
\address[3]{School of Information Technology, Deakin University, 221 Burwood Highway, VIC 3125, Australia.}
\address[4]{School of Computing,
Newcastle University, Newcastle upon Tyne NE1 7RU, United Kingdom.}


\markboth
{Naha \headeretal: Fog Computing: Survey of Trends, Architectures, Requirements, and Research Directions}
{Naha \headeretal: Fog Computing: Survey of Trends, Architectures, Requirements, and Research Directions}

\corresp{Corresponding author: Longxiang Gao (e-mail: longxiang.gao@deakin.edu.au).}

\begin{abstract}

Emerging technologies like the Internet of Things (IoT) require latency-aware computation for real-time application processing. In IoT environments, connected things generate a huge amount of data, which are generally referred to as big data. Data generated from IoT devices are generally processed in a cloud infrastructure because of the on-demand services and scalability features of the cloud computing paradigm. However, processing IoT application requests on the cloud exclusively is not an efficient solution for some IoT applications, especially time-sensitive ones. To address this issue, Fog computing, which resides in between cloud and IoT devices, was proposed. In general, in the Fog computing environment, IoT devices are connected to Fog devices. These Fog devices are located in close proximity to users and are responsible for intermediate computation and storage. One of the key challenges in running IoT applications in a Fog computing environment are resource allocation and task scheduling. Fog computing research is still in its infancy, and taxonomy-based investigation into the requirements of Fog infrastructure, platform, and applications mapped to current research is still required. This survey will help the industry and research community synthesize and identify the requirements for Fog computing. 
This paper starts with an overview of Fog computing in which the definition of Fog computing, research trends, and the technical differences between Fog and cloud are reviewed. Then, we investigate numerous proposed Fog computing architecture and describe the components of these architectures in detail. From this, the role of each component will be defined, which will help in the deployment of Fog computing. Next, a taxonomy of Fog computing is proposed by considering the requirements of the Fog computing paradigm. We also discuss existing research works and gaps in resource allocation and scheduling, fault tolerance, simulation tools, and Fog-based microservices. Finally, by addressing the limitations of current research works, we present some open issues, which will determine the future research direction for the Fog computing paradigm.

\end{abstract}



\titlepgskip=-15pt

\maketitle

\section{Introduction}
\label{sec:introduction}
\PARstart{I}{ndividuals} and organizations are increasingly becoming dependent on computers and smart devices to deal with daily tasks. These devices are generating data via various sensors and applications. As a result, organizations are generating and storing huge amounts of data on a regular basis \cite{assunccao2015big}. After the proliferation of IoT, data generated by sensors has increased enormously. With this sudden increase in the volume of data being produced and inability of conventional databases to process various forms of structured and unstructured data, big data analytics has attained great attention in recent years. Every organization is now prioritizing the analysis of collected data to extract useful insights in order to make important decisions \cite{chen2013big}. Nowadays, organizations need a dynamic IT infrastructure because of the shift to cloud computing due to its accessibility, scalability, and pay-per-use features. The most common services provided by the cloud are known as Software as a Service (SaaS), Platform as a Service (PaaS), and Infrastructure as a Service (IaaS), all of which are heading towards Anything as a Service (XaaS) \cite{alhaddadin2014user}. However, data generated from billions of sensors, referred to as big data, cannot be transferred and processed in the cloud. In addition, some IoT applications need to be processed faster than the cloud's current capability. This problem can be solved by using the Fog computing paradigm, which harnesses the processing power of devices located near users (idle computing power) to support utilization of storage, processing, and networking at the edge \cite{dastjerdi2016Fog}. 

Fog computing is a decentralized computing concept, which does not exclusively rely on any central component like cloud computing \cite{mahmud2016Fog,gao2017Fogroute}. It is able to overcome the high latency problem of the cloud by using idle resources of various devices near users. However, Fog computing relies on the cloud to do complex processing. Unlike cloud computing, Fog computing is a decentralized computing concept, where the many devices around us, which have computation capacity, are utilized. Currently, even a low- specification smartphone has processing capacity, sometimes with multiple cores. Hence, many devices like smartphones, switches, routers, base stations, and other network management devices equipped with processing power and storage capacity can act as Fog devices. The resources of these devices are idle outside of peak hours.

Many research issues relating to Fog computing are emerging due to its ubiquitous connectivity and heterogeneous organization. In the Fog computing paradigm, key issues are the requirements and the deployment of Fog computing environment. This is because the devices that exist in Fog environments are heterogeneous: therefore, the question that arises is how will Fog computing tackle the new challenges of resource management and failure handling in such a heterogeneous environment? Hence, it is necessary to investigate the very basic requirements for all other related aspects such as deployment issues, simulations, resource management, fault tolerance, and services. Several reviews \cite{yi2015survey,hu2017survey,baccarelli2017Fog,varshney2017demystifying,perera2017fog,mahmud2018Fog,mouradian2018comprehensive} have been done on Fog computing. Here, we present the focus and survey domains of these review works in brief.

Similar concepts of Fog computing, definitions, application scenarios, and numerous issues are described by one study \cite{yi2015survey}.
Hu et al. \cite{hu2017survey} presented the hierarchical architecture of Fog computing and technologies like computing, communication, and storage technologies, namely resource management, security, and privacy protection that support Fog deployment and application. Baccarelli et al. \cite{baccarelli2017Fog} surveyed Fog computing and the Internet of Everything (IoE) with an integrated point of view of Fog computing and IoE. Varshney et al. \cite{varshney2017demystifying} reviewed various dimensions of application characteristics, system architecture, and platform abstractions of edge, Fog, and cloud ecosystems. Perera et al. \cite{perera2017fog} reviewed the Fog computing domain from the platform perspectives of developers and end users towards building a sustainable sensing infrastructure for smart city applications. Mahmud et al. \cite{mahmud2018Fog} presented a taxonomy of Fog computing according to the identified challenges and its key features. The proposed taxonomy provides a classification of the existing works in Fog computing. Mouradian et al. \cite{mouradian2018comprehensive} reviewed Fog architecture and algorithms based on six different evaluation criteria, namely heterogeneity, QoS management, scalability, mobility, federation, and interoperability. However, none of the studies had investigated taxonomy based on the requirements of infrastructure, platform, and application in Fog computing. Moreover, none of them comprehensively investigated fault tolerance, resource management, or microservices in Fog computing. We consider the aforementioned current issues and discuss these extensively and also highlight how cloud computing-related solutions could be employed in the Fog in some cases. The contributions of this review work can be summarized as follows:

\begin{itemize} 
\item Present the research trends in Fog computing by investigating the number of published research works and search occurrences in Google Scholar. 
\item Review of several Fog computing architectures and presentation of a detailed architecture, as most of the previous researchers only presented high-level architecture.
\item	Present a taxonomy by considering the requirements of infrastructure, platform, and application in the Fog computing paradigm.
\item Identify Fog computing research gaps in resource allocation and scheduling, fault tolerance, simulation tools, and Fog-based microservices.
\item Address the limitations of current research works and some open issues in infrastructure, platform, and applications. 
\end{itemize}

From this survey, the industry and research community will be able to gain insight into the requirements for building a Fog computing environment with a better understanding of resource management in the Fog.

The remainder of this paper is organized as follows: Section II surveys definitions and research trends in Fog computing with a technical comparison between Fog and cloud paradigms. Section III discusses computing paradigms similar to Fog computing. Section IV presents related works on Fog computing architecture and discusses the components of the Fog computing architecture. Section V shows the taxonomy of Fog computing by reviewing its requirements. Section VI presents various application dimension of Fog computing. Section VII discusses current state-of-the-art Fog computing technology. Section VIII presents open issues and future research direction. Section IX concludes the paper.
   
\section{Overview of Fog computing}

The term `Fog computing' was proposed in 2012 by researchers from Cisco Systems \cite{bonomi2012Fog}. Processing application logic and data at the edge is not a new concept. The concept of Edge computation emerged around the 2000s \cite{xie2002enabling, gelsinger2001microprocessors} and another similar concept, cloudlets, was introduced in 2009 \cite{ibrahim2009cloudlet}. Both Cloudlets and Fog computing are the advancements of a similar concept, which revolves around processing at the edge level. While cloudlets are applied in the mobile network, Fog computing is applied to connected things such as IoT, which plays into the concept of IoT \cite{gonzalez2016Fog}.

Fog is both a virtualized and non-virtualized computing paradigm that provides networking, storage, and computation services amid cloud servers and IoT devices \cite{bonomi2012Fog,dastjerdi2016Fog}. However, these services are not completely located at the network edge. The Fog is a distributed computing approach that mainly focuses on facilitating applications, which require low latency services \cite{li2017latency}, Fog computing also supports non-latency aware services. It is obvious that using idle computation resources near the users will improve overall service performance, if the volume of processing were not that high. A huge number of heterogeneous nodes will be connected to the Fog. These nodes include sensors and actuators among others  \cite{bonomi2012Fog}. Computation is performed in Fog devices when necessary and storage facilities are also available for a short period of time, at least in most Fog devices. Time-sensitive computation in the Fog is done without the involvement of third parties, and in most cases, is done by the Fog processing devices. According to Yi et al. \cite{yi2015survey}, the Fog computing paradigm supports the running of new services or basic network functions and applications in a sandboxed environment similar to cloudlets. However, the subject is still a research challenge because the question of how the Fog will provide these service still remains. In addition, will the Fog have cloud service providers or will it be like a single entity as a whole? Figure \ref{fig_Fogcom} shows a basic model of Fog Computing. Fog devices, Fog servers, and gateways are the basic computation components in the Fog environment. Any device that has computation, networking, and storage capabilities can act as a Fog device. These devices include set-top boxes, switches, routers, base stations, proxy servers or any other computing device. Fog servers that manage several Fog devices and Fog gateways are responsible for translation services between heterogeneous devices in the Fog computing environment. Fog gateways also provide translation services between IoT, Fog, and cloud layers. New challenges in this emerging computing paradigm have emerged in the past couple of years.

In this section, we discuss the various definitions of Fog computing and define Fog computing from our point of view. In addition, we discuss and analyze research trends in Fog computing. Finally, we compare the technical differences between Fog computing and cloud computing.

\Figure[t!](topskip=0pt, botskip=0pt, midskip=10pt)[width=3in]{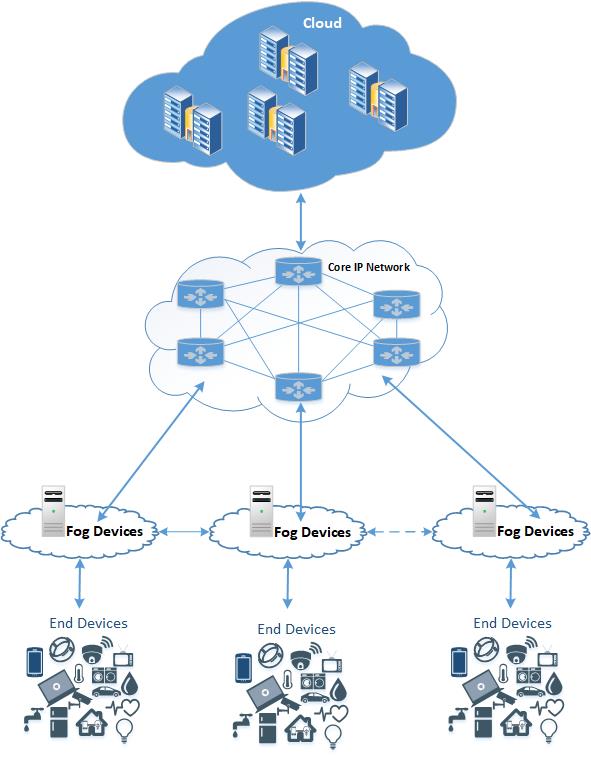} {A model  of Fog computing.\label{fig_Fogcom}}


\subsection{Definition of Fog computing}

Fog computing is a distributed computing paradigm where processing is done at the edge of the network with seamless integration of the cloud infrastructure. It enables a computing facility for IoT environments or other latency sensitive application environments. It is estimated that about 50 billion ``things'' will be connected to the Internet by 2020 \cite{cisco2015Fog}. Transferring all data from all connected devices for processing on the cloud will need massive amounts of bandwidth and storage. All devices are not connected to the controller via IP but connected by some other IoT industrial protocols. Because of this, a translation process is also needed for the processing or storing of information from IoT devices. Various researchers have defined Fog computing in different ways. Some examples are as follows:

\begin{itemize}
	\item \textit{``Fog computing is a highly virtualized platform that provides compute, storage, and networking services between IoT devices and traditional cloud computing data centers, typically, but not exclusively located at the edge of network.'' \cite{bonomi2012Fog}}
	\item \textit{``Fog computing is a scenario where a huge number of heterogeneous (wireless and sometimes autonomous) ubiquitous and decentralised devices communicate and potentially cooperate among them and with the network to perform storage and processing tasks without the intervention of third parties. These tasks can be for supporting basic network functions or new services and applications that run in a sandboxed environment. Users leasing part of their devices to host these services get incentives for doing so.''\cite{vaquero2014finding}} 
	\item \textit{``The term Fog computing or Edge Computing means that rather than hosting and working from a centralized cloud, Fog systems operate on network ends. It is a term for placing some processes and resources at the edge of the cloud, instead of establishing channels for cloud storage and utilization.'' \cite{IBMcloudcomputingnews_2016}}	
\end{itemize}

The first definition of Fog computing was presented by Bonomi et al. \cite{bonomi2012Fog}, where they addressed the computing paradigm as a highly virtualized platform. However, some IoT devices such as smartphones are not virtualized but could also be a part of the Fog infrastructure, as some processing could still be done. According to Cisco \cite{CiscoSystems2016b}, the Fog computing paradigm provides an ideal place to analyze most data near the devices that produce and act on that data instantaneously. The Fog is located near things that are able to process and act on the data generated. The devices that are within the Fog environment are known as Fog devices. These nodes can be deployed at any place with a connectivity to the network: on the power pole, on the factory floor, alongside the road, alongside the railway line, in a vehicle, inside a shopping mall, on an oil rig, etc. A device that has processing, storage, memory, and network capability can act as a Fog device. Although the Fog extends the cloud, technically it resides in between the cloud and IoT devices and handles processing and storage tasks in close proximity to the user. Yi et al. \cite{yi2015survey} stated that the definition given by Vaquero and Rodero-Merino \cite{vaquero2014finding} is debatable and a definition that can distinguish clearly between Fog computing and other related computing paradigms is still required. The definition given by IBM \cite{IBMcloudcomputingnews_2016} represents Edge and Fog computing as the same computing paradigm. According to Shi et al. \cite{shi2016edge}, Fog computing focuses more on the infrastructure side while edge computing focuses more on the things' side.  Furthermore, Edge computing is not spontaneously associated with any cloud-based services such as SaaS, IaaS, and PaaS \cite{mahmud2016Fog}. In brief, Table \ref{Figdef} summarizes Fog definitions provided by various research works. 

\begin{table}[!t]
	\centering
    \small
	\caption{Summary of Fog computing definitions}
	\label{Figdef}
	\begin{tabular}{L{2cm}L{5cm}} 
		\toprule
		\textbf{Defined by} & \textbf{Characteristics} \\ \midrule
		Bonomi et al. \cite{bonomi2012Fog} &  Highly virtualized \\
		\cmidrule{2-2}  &  Reside between IoT devices and cloud \\
		\cmidrule{2-2}  &  Not exclusively located at the edge \\ \midrule
		Cisco Systems \cite{CiscoSystems2016b}  & Extends the Cloud \\  
		\cmidrule{2-2}  & Generally used for IoT \\
		\cmidrule{2-2}  &  Can be deployed anywhere\\
		\cmidrule{2-2}  &  Fog device consists of processing, storage, and network connectivity\\ \midrule
		Vaquero and Rodero-Merino \cite{vaquero2014finding}  & heterogeneous, ubiquitous and decentralised devices communication \\
		\cmidrule{2-2}  &  Storage and processing done  without third party invention \\
		\cmidrule{2-2}  &  Run in a sandboxed environment \\
		\cmidrule{2-2}  &  Leasing part of users devices and provide incentive \\ \midrule
		IBM \cite{IBMcloudcomputingnews_2016} & Defined Fog and Edge computing as similar \\
		\cmidrule{2-2}  &  Not depends on centralized cloud \\
		\cmidrule{2-2}  &  Resides at network ends \\
		\cmidrule{2-2}  &  Place some resource and at the edge of the cloud\\ \midrule
		Proposed Definition & Virtualization and non-virtualization characteristics \\
		\cmidrule{2-2}  &  Association with the cloud for non-latency-aware processing and storage \\
		\cmidrule{2-2}  &  Any edge device with available processing power and storage capability can be act as a Fog device \\
		\cmidrule{2-2}  &  Always resides between end users and cloud \\ \midrule
		\bottomrule
	\end{tabular}
\end{table}

Considering the above definitions, we define Fog computing as follows:

\begin{itemize}
    \item
 \textit{Fog computing is a distributed computing platform where most of the processing will be done by virtualized and non-virtualized end or edge devices. It is also associated with the cloud for non-latency-aware processing and long-term storage of useful data by residing in between users and the cloud}. \end{itemize}

In our definition, we considered all devices with computing and storage capacity as Fog devices and also more precisely identified the role of the cloud in the Fog computing environment.

\subsection{Fog computing research trends}

Growing attention towards processing data closer to the users has been observed among industries and the academia in the past few years. Handling IoT-generated data at the edge level will help improve overall processing time. In this section, we investigate Fog and other related technological trends for the past few years in the research community. According to the Gartner hype cycle, in July 2017 \cite{gartnerhype}, the peak emerging technology is the smart home, which would perform better with the incorporation of the Fog computing environment. A Hype Cycle \cite{gartnerhype} represents common patterns of new trending technologies. Fog computing can also enable latency-aware smart home services in a more efficient and convenient way, especially for emergency response smart home applications. According to the Gartner hype cycle demonstration, some other influencing technologies include virtual assistants, autonomous vehicles, IoT platforms, smart robots, edge computing, and smart workspaces, which are required to support latency-aware applications. All these mentioned technologies could benefit from the support of the Fog computing paradigm due to latency sensitiveness, connectivity to the cloud, and edge-level data processing capability. Except for the autonomous vehicle technology, all other aforementioned technologies will reach the market adoption threshold in the next 10 years. Besides the hype cycle analysis, we analyzed the search occurrence of Fog and other related technologies in Google Scholar. In addition, the number of papers available in different digital libraries related to the Fog was also analyzed.  



Google Scholar search occurrences of various similar technologies to Fog were investigated in the past few years, as presented in Figure \ref{gsc12}. According to the data, edge computing is the topmost searched item in Google Scholar compared to other similar technologies. However, the search trend decreased by more than three times in the past eight years for edge computing. Mobile cloud computing and mobile edge computing are the other two top-searched computing paradigm after edge computing. The lowest trend observed was for dew computing and Fog dew computing. While the trend for edge computing is decreasing, Fog computing related to scholarly searches is increasing year by year, and has increased by 2.5 times from 2010 to 2017. This shows that Fog computing is the fastest growing research area in academia and will have a great impact on the industry as well.

\Figure[t!](topskip=0pt, botskip=0pt, midskip=10pt)[width=6in]{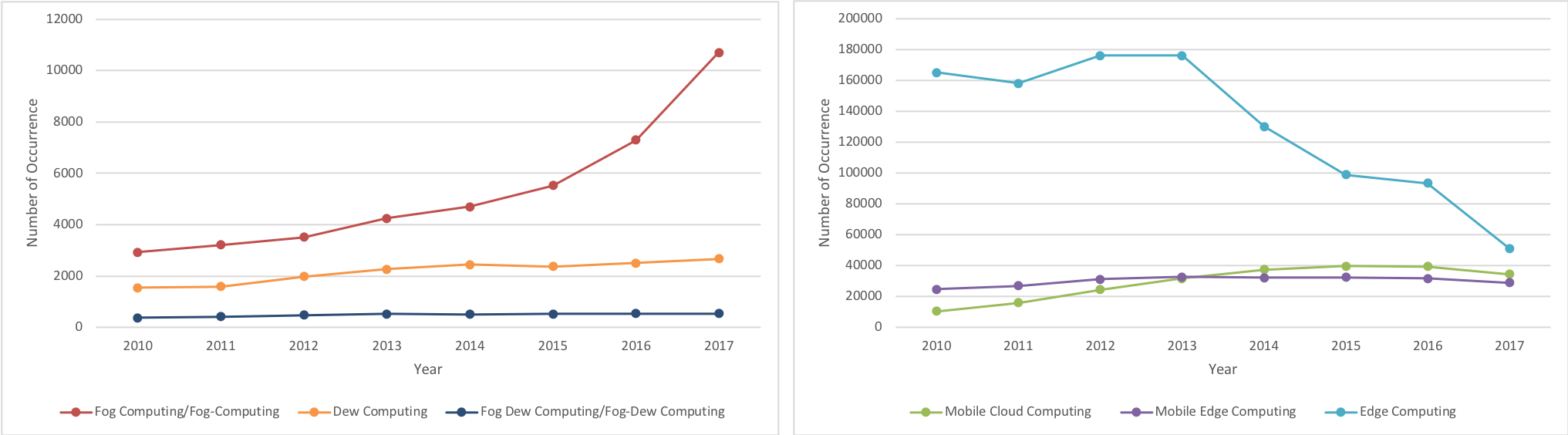} {Search occurrence of similar technologies like the Fog in Google Scholar.\label{gsc12}}


Fog computing topic search in the Web of Science shows that the number of scholarly articles has more than doubled between 2015 and 2016, as per Figure \ref{wosFog}. The first paper with `Fog computing' in its title was published in 2012. Since then, about 564 journal and conference articles have been published on this topic in the four major digital libraries (Web of Science, Science Direct, IEEE Xplore, and ACM), as presented in Figure \ref{4dlsearchFog}. Cloud computing first emerged in 2008 \cite{buyya2009cloud}. This shows that Fog computing publications have dramatically increased, as no study in this area was seen in the couple of years following the introduction of cloud computing research in 2008 (see Figure \ref{cloudwostitlesearch}).

\Figure[t!](topskip=0pt, botskip=0pt, midskip=10pt)[width=3in]{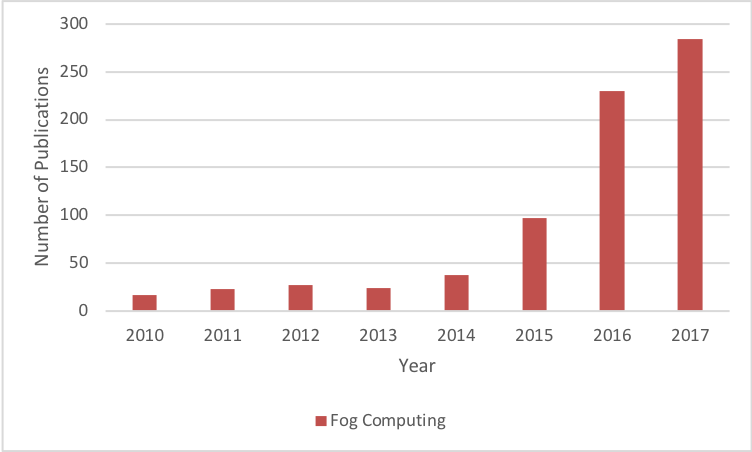} {No. of Fog computing-related papers in the Web of Science (as Feb 2018).\label{wosFog}}


\Figure[t!](topskip=0pt, botskip=0pt, midskip=10pt)[width=3in]{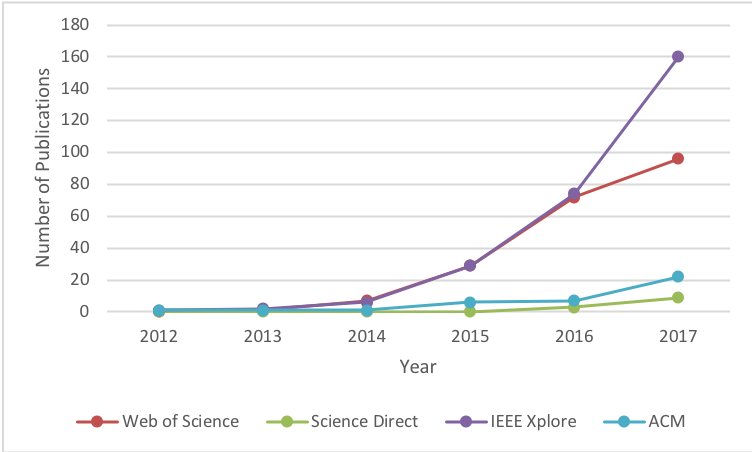} {Number of publications with ``Fog computing" in the title in the four major digital libraries.\label{4dlsearchFog}}


\Figure[t!](topskip=0pt, botskip=0pt, midskip=10pt)[width=3in]{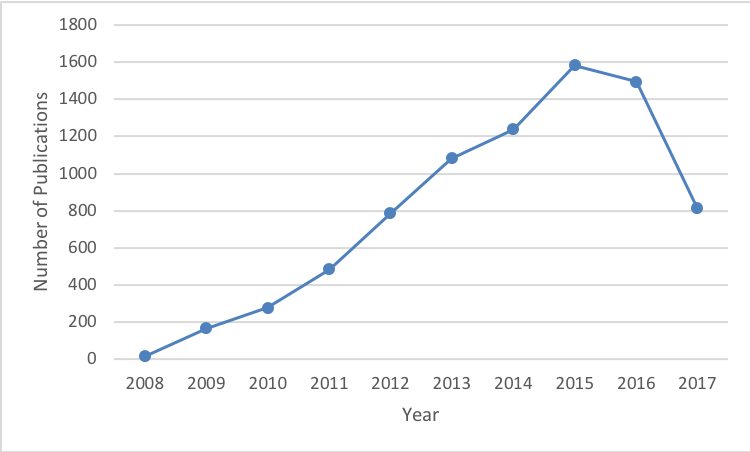} {Published articles with the title cloud computing in the Web of Science.\label{cloudwostitlesearch}}


From our observation, it is obvious that the interest in Fog computing research is rapidly increasing. Idle resources in the form of devices near users can be utilized within the Fog computing concept. Thus, a clear direction to market the adoption and technological development of Fog deployment has emerged.

\subsection{Difference between Fog and cloud}
Fog computing architectures are based on Fog clusters where multiple Fog devices participate to cooperate with the processing. On the other hand, datacenters are the main physical components of clouds. Because of this, cloud computing has high operational costs and energy consumption. By comparison, energy consumption and operation costs in the Fog computing paradigm is low. The Fog is located closer to the user, so the distance between users and Fog devices could be one or a few hops, which is also agreed by Hu et al. \cite{hu2017survey}. However, according to Mahmud et al. \cite{Mahmud2018cloudFog}, the distance between users to the Fog is one or two hops. Again, Luan et al. \cite{luan2015Fog} argued that the distance should be one hop with wireless connectivity. Yet, all agreed with the distance between the users to the cloud, which is a multi-hop distance. Due to the distance, communication latency for the cloud is always high compared to the Fog. The cloud is a more centralized approach while the Fog is a more distributed approach based on geographical orchestration \cite{Mahmud2018cloudFog}. 

Real-time Interaction is not possible for the cloud due to its high latency, but this problem can be easily resolved by Fog computing. On the other hand, the rate of failure in the Fog is high because of wireless connectivity, decentralized management, and power failure \cite{wang2015Fog,Mahmud2018cloudFog,syed2016pattern,yi2015Fog}. Most devices in Fog environments will be connected wirelessly since smart gadgets and handheld devices will be participating in Fog systems \cite{chiang2016Fog}. These devices, and other network management devices, are mostly decentralized. These devices could fail when software is not managed correctly. Users may not be aware of malicious software that could lead to device failure. Moreover, Fog processing could fail in other cases as well, for example, each Fog device is responsible for performing its own application processing. So, the IoT application processing in a Fog device always takes second priority. If the Fog device is fully utilized by the application of the device itself, then it will fail to do any Fog processing. Hence, the scheduling of applications and resources in the Fog is more complex. In addition, failure handling in the Fog is competitive because of power failure, which is only an issue because the devices run on battery power. Altogether, Table \ref{cloudFogdif} shows the technical differences between the cloud and the Fog.

\begin{table*}[ht]
	\centering
    \small
	\caption{Technical difference between Fog and cloud}
	\label{cloudFogdif}
	\begin{tabular}{L{4cm}L{4cm}L{4cm}} 
		\toprule
		& \textbf{Fog} & \textbf{Cloud} \\ \midrule
        Participating Nodes	& Constantly dynamic	& Variable \\ \midrule
        Management &	Distributed/Centralized	& Centralized \\ \midrule
        Computation device	& Any device with computation power	& Powerful Server System \\ \midrule
        Nature of Failure	& Highly diverse	& Predictable \\ \midrule
        Connectivity from user	& Mostly wireless	& High speed (With the combination of wire and wireless) \\ \midrule
        Internal connectivity &	Mostly wireless	& Mostly Wired \\ \midrule
        Power source	& Battery/Direct power/Green Energy, such as solar power	& Direct power \\ \midrule
        Power consumption	& Low	& High \\ \midrule
        Computation capacity	& Low	& High \\ \midrule
        Storage capacity	& Low	& High \\ \midrule
        Network latency	& Low	& High \\ \midrule
        Node mobility	& High	& Very low \\ \midrule
        Number of intermediate hop	& One/Few	& Multi \\ \midrule
        Application type	& latency-aware	& Non latency-aware \\ \midrule
        Real time application handling	& Achievable	& Difficult \\ \midrule
        Computation cost	& Low	& High \\ \midrule
        Cooling cost	& Very low	& High \\ \midrule
        Space required for deployment &	Very little, also possible to install at outdoor on existing infrastructure	& Warehouse size building \\ \midrule
		\bottomrule
	\end{tabular}
\end{table*}

Definitely, it cannot be said that the Fog can replace the cloud. We cannot even conclude that the Fog is better than the cloud either, both contribute differently via fulfilling different perspectives and requirements.

\section{Related paradigms and technologies}

Fog computing uses computing resources near underlying networks, located between the traditional cloud and edge devices, to provide better and faster application processing and services \cite{bonomi2012Fog}. Several similar computing paradigms exist besides Fog computing such as Mobile Cloud Computing (MCC), Mobile-Edge Computing (MEC), Edge Computing, Dew Computing, and Fog-dew computing. In cloud computing, all IoT devices are directly connected to the cloud and computation totally depends on the cloud. However, all the above similar technologies do not exclusively depend on the cloud, but depend on some intermediate devices for computation; some of them do not even require a connection to the cloud. Figure \ref{relFog} shows the high-level architecture of these technologies.

\Figure[t!](topskip=0pt, botskip=0pt, midskip=0pt)[width=3in]{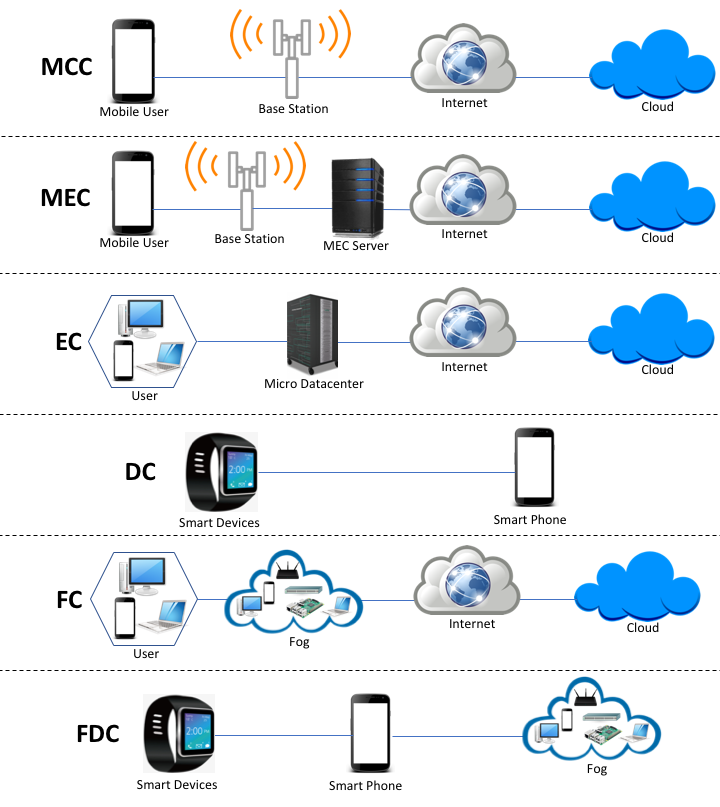} {High level architecture of Mobile Cloud Computing (MCC), Mobile Edge Computing (MEC), Edge Computing (EC), Dew Computing (DC), Fog Computing (FC) and Fog Dew Computing (FDC)\label{relFog}}


\subsection{Mobile Cloud Computing (MCC)}
Remote execution of offloaded mobile services is done with the support of MCC near end users \cite{sanaei2014heterogeneity,bahl2012advancing}. MCC overcomes the computational, energy, and storage resource limitation of smart mobile devices. Generally, a lightweight cloud server (cloudlet) is placed at the edge of the network \cite{satyanarayanan2013role} to overcome these issues. MCC is a mobile computing technology, which provides unrestricted functionality, mobility, and storage facility through heterogeneous network connectivity. This technology also provides unified elastic computing resources by following the pay-per-use model. It also provides access to data, application, and cloud via the Internet for mobile users. It is expected that this technology will be applied in education, urban and rural development, healthcare, and more realistic social networking in the future \cite{sanaei2014heterogeneity}. Nowadays, many computation-intensive applications are widely available, such as Augmented Reality, computer vision and graphics, speech recognition, machine learning, planning and decision-making, and natural language processing applications. However, simply designing powerful mobile devices will not meet the requirements for these applications \cite{bahl2012advancing}. Rather, the applications require edge processing as well as collaboration with the cloud for complex processing. Thus, mobile computing demands fundamental changes to cloud computing, for example, a low-latency middle tier, programming models to enable seamless remote execution, basic mobile cloud services such as presence services, cloud infrastructure optimization for mobile applications, memcache services, and so on \cite{bahl2012advancing}. The convergence of mobile cloud computing is predicated on a reliable, end-to-end network, and high bandwidth, which isdifficult to guarantee in harsh environments. One of the solutions to this deep-rooted problem is the VM-based cloudlets located at a closer location to the mobile device \cite{satyanarayanan2013role}.

\subsection{Mobile Edge Computing (MEC)}
MEC proposes the co-location of computing and storage resources at the base stations of cellular networks \cite{Beck2014a}. MEC could either be connected or disconnected to cloud datacenters in a remote location. Hence, MEC supports two- or three-tier hierarchical application deployments along with end mobile devices \cite{klas2015Fog}. 
In a MEC ecosystem, a new device called the MEC server needs to be deployed near base station towers to provide processing and storage capabilities at the edge. Four participants are involved in this computing paradigm, which are the mobile end users, network operators, Internet infrastructure provider (InPs),and application service provider. Mobile end users are the main consumer of the system and request their service via user equipment (UE). Network operators manage and maintain the operation of base stations, mobile core network, and MEC servers. InPs maintain Internet connectivity and routers. Application service providers host the application services in the content delivery networks (CDN) or within a data centers. Processing of requests from the UE will search out the closest MEC. The MEC server is capable of processing user request instead of forwarding it to remote Internet services. In a case where it is not possible to process or complete a request  at the MEC sever; the request will be forwarded to remote CDNs or data centers \cite{Beck2014a}.

According to Klas \cite{klas2015Fog}, MEC is the evolution of mobile base stations. It is a natural development. It is a collaborative deployment of telecommunication and IT networking. This computing paradigm enables new vertical business segments and services to individual end users and enterprise consumers. Various services could be delivered through this computing paradigm including IoT, location services, augmented reality, caching service, video analytics, and local content distribution. It can deliver real-time low-latency access of local content or by caching content at the MEC server. However, the main limitation of this system is the installation of the MEC server, which is specifically dedicated to MEC services. Scaling is another big issue with the increase in resource demand over time.

\subsection{Edge computing}
Edge devices or edge servers provide computation facilities in Edge computing. In general, edge computing does not spontaneously associate with any types of cloud-based services and concentrates more on the IoT device side \cite{shi2016edge}. 
One study defined the edge as any network or computing resource near the path between cloud data centers and data sources \cite{shi2016edge}. Any smart device or sensor could have data sources but the edge is different. For example, a cloudlet and a micro datacenter is the edge of the mobile application and cloud, whereas the IoT gateway is the edge between IoT sensors and cloud. Similarly, if a cloud application is running on a smartphone, then the smartphone is the edge of the application and the cloud  \cite{satyanarayanan2009case}. The main motivation of edge computing is that the computation should be done at a closer location to the data sources.

In the edge computing concept, things are not only consuming data but also produce data by taking part in processing. Edge devices can perform computation task from the cloud besides requesting services and content. Data storage, computing offloading, processing, and caching will be done by an edge node. The edge device is also capable of distributing requests and providing service on behalf of the cloud to the users. In such scenarios, edge devices need to be well designed to meet privacy requirements, reliability measures, and security concerns \cite{shi2016edge}.

\subsection{Dew Computing (DC)}
In the current computing hierarchy, Dew Computing \cite{Wang2015} is situated at the ground level of the cloud and Fog computing environment \cite{skala2015scalable}. DC goes beyond the concept of service, storage, and network, to a sub-platform, which is based on a microservice concept for which its computing hierarchy is vertically distributed  \cite{skala2015scalable}. The DC approach facilitates resources such as sensors, tablets, and smartphones that are seamlessly connected to a network. Because of this, DC covers a wide range of ad-hoc-based networking technologies \cite{skala2015scalable}. 

Skala et al. \cite{skala2015scalable} argued that DC is much more useful in everyday life compared to Fog computing. Fog supports IoT-based applications, which demand less latency and real-time capability and a dynamic network configuration while DC is microservice concept and thus is not dependent on any centralized device, server, or cloud. They provide an example in which DC could be integrated into a smart traffic control system, where data collection and processing units will be located in between the traffic signals. These units generate a collective overview of the current traffic conditions. In such a way, a car with low fuel will be notified before entering heavy congestion, or a hybrid car will be informed of switching to conventional fuel before approaching the congestion. As a result, cars with less fuel will be rescued from unwanted situations and hybrid cars could reduce exhaust smoke densities significantly. Although the concept is microservice-based, the processing is completed in Fog devices. In the Fog computing concept, it is not crucial that applications must be dependent on the cloud or require the storing of results in the cloud. On the other hand, if such traffic processing information were stored, it would help strategic decision-making to improve traffic management. Dew computing is an emerging research area and its goal is to use the full potential of cloud and local resources \cite{Wang2015a}.

\subsection{Fog-Dew computing}
In the architecture of Fog-dew computing, IoT devices need not have an active Internet connection while being connected to the community server. The community server will interact with the cloud and is responsible for providing services to the IoT devices \cite{mane2017addincondewFog}.

Cloud computing always needs an Internet connection, which is the main drawback of the cloud. While the cloud is unable to serve users without an Internet connection, Fog-dew computing facilitates offline services without an Internet connection. However, there are some exceptions. For example, the navigation app, Waze, allows users to navigate offline. This feature was also recently added to Google Maps. In this case, a map information file for a specific area is downloaded to the user device and allows users to navigate during an offline state. Another example is Google Drive and Dropbox, where users can delete, create, and update files and folders in offline mode and then sync once the device is connected to the Internet. However, these services are not purely offline-we may not rely on the Internet directly but we cannot completely ignore Internet connection. The situation becomes more complex when a single user uses multiple offline devices alongside the complexity that arises in a multiuser environment. Such situations could be mitigated with the help of Fog-dew computing.

In the Fog computing paradigm, IoT devices are connected to the cloud via Fog devices. Fog devices are connected to the cloud through the core network. Fog computing is a combination of MEC and MCC \cite{yi2015survey} but the main goal of all Fog-related paradigms is to perform processing at the edge. These related paradigms differ from each other based on Internet and cloud connectivity. Also, the amount of processing that needs to be done at the edge differs based on service requirements. Furthermore, the type of devices that will be used for computation and storage purposes is also another issue. In summary, Table \ref{rttable} shows the characteristics of the above-discussed related computing paradigms along with the Fog computing paradigm.

{\renewcommand{\arraystretch}{1.3}
\begin{table*}[htbp]
	\centering
    \scriptsize
	\caption{Summary of similar technologies like Fog}
	\label{rttable}
	\begin{tabular}
    {L{2cm}|C{1.5cm}|C{1.5cm}|C{1.2cm}|C{1.5cm}|C{1.5cm}|C{3.5cm}} \hline \toprule
		\multirow{3}{*} \textbf{Computing Paradigm \& Applications}&\textbf{MCC}&\textbf{MEC}&\textbf{EC}&\textbf{DC}&\textbf{Fog-DC}&\textbf{Fog}\\ \cline{2-7}
		
		&\multicolumn{6}{|c}{Offload mobile applications to the computation unit closer to the user}\\ \cline{2-7}
        
         &Mobile apps&Mobile apps& IoT related apps & Smart Devices (Fitbit, health monitoring)& Smart Devices (Fitbit, health monitoring)& IoT apps, Mobile apps, Video streaming, smart grid, smart transportation system, big data processing, stream processing\\ \hline
        
        \textbf{Connection to the Cloud}& Yes & Yes or No & Yes or No & No & No & Yes	\\ \hline
        
        \textbf{Types of Users}& Mobile & Mobile & Mobile / Stationary & Smart sensor based devices & Smart sensor based devices & Mobile / Stationary	\\ \hline
        
        \textbf{Virtualization Technology}& Hypervisor / Container & Hypervisor / Container & Hypervisor / Container & Container & Container & Hypervisor / Container	\\ \hline
        
        \textbf{Main Computation Element}& Base station server & MEC server & Micro Data Center & Smartphone & Smartphone & Any device with the capabilities of computation, storage, memory and network adapter \\ \hline
        
        \textbf{Example of Commercial Prototype}& Akamai \cite{akamai2018} & - & Ec-IoT (Huawei) \cite{eciot2018} & - & - & IOx (Cisco) \cite{cisco_2016IOx}	\\ \hline
        
        \textbf{Example of  R\&D Prototype}& - & Hyrax \cite{hyrax2018}, Saguna Open-RAN \cite{saguna2018} & -& - & - & -	\\ \hline
        
        \bottomrule\end{tabular}
\end{table*}
}


\section{Architecture of Fog computing}
For market adoption and deployment, Fog computing must have a standard architecture. There is no available standard architecture to date. However, many research works have presented Fog computing architectures. In this section, we firstly discuss the high-level architecture of Fog computing. Furthermore, we summarize some proposed architectures for Fog computing. Finally, we present a detailed architecture for Fog computing with a comprehensive description of each component of the architecture.  

\subsection{ High-level architecture of Fog computing}
In high-level architecture, the Fog computing paradigm has three different layers, as shown in Figure \ref{hlarc}. The most important layer is the Fog layer. This layer consists of all intermediate computing devices. Traditional virtualization technologies can be used at this plane, similar to the cloud. However, considering the resource availability, employing container-based virtualization is more appropriate. This layer accumulates sensor-generated data from the IoT layer and sends an actuation-related request after processing. Although it seems that the big data problem is solved by processing generated data at the edge level, billions of devices will create big data issue. In fact, it is possible to employ small- and medium-scale big data processing at this level. Many research works have been undertaken to process big data in the Fog plane  \cite{ dubey2015Fog, Ahmad2016, tang2017incorporating, tang2015hierarchical, Zhang2017, Yin2017, Pecori2018}.

\Figure[t!](topskip=0pt, botskip=0pt, midskip=0pt)[width=3in]{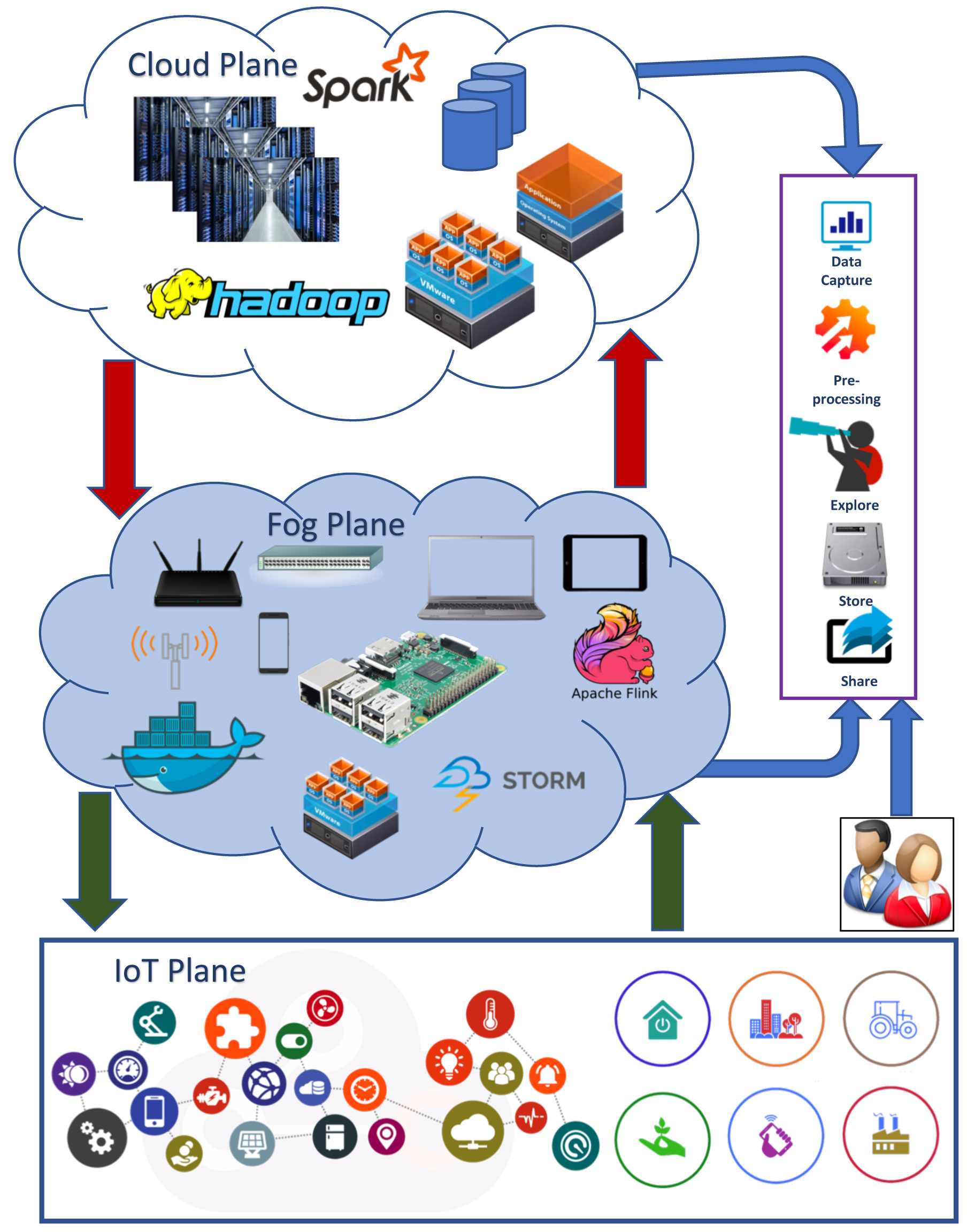} {High level Fog computing architecture.\label{hlarc}}


The bottommost layer is the IoT plane, which consists of all connected devices. The devices on this plane perform the sensing and actuation process. For time-sensitive applications, processing should be done on the Fog plane exclusively while the cloud can perform other processing that is not time-sensitive. However, the Fog layer will manage what needs to be sent to the cloud and what should not. The users are able to get services from both the Fog and cloud based on their request. However, the cloud plane will manage complex processing and storage.

\subsection{Various proposed architectures for Fog computing}
Layered representation is the best way to represent Fog architecture. Many works have been done to quantify the layer-based concept of Fog architecture \cite{aazam2015Fog, arkian2017mist, luan2015Fog, dastjerdi2016Fog, nadeem2016Fog, taneja2016resource, sarkar2016theoretical}. From our review, we found that researchers have proposed three \cite{luan2015Fog, nadeem2016Fog, taneja2016resource, sarkar2016theoretical}, four \cite{arkian2017mist}, five \cite{dastjerdi2016Fog}, and six \cite{ aazam2015Fog} layers in the Fog architecture. 

Everyone has their own justifications for their claims. If we ignored the user plane, it is obvious that Fog architecture could be defined as three different levels from the high level. As we proceed to the more implementation-type level, the number of layers in the architecture would vary, giving rise to five \cite{dastjerdi2016Fog} and six \cite{ aazam2015Fog} levels in the Fog computing layer. 

Aazam and Huh \cite{ aazam2015Fog} presented six different layers based on specific tasks. On the other hand, Dastjerdi et al. \cite{dastjerdi2016Fog} defined five different layers based on network perspective. Other high-level architectures in Fog computing were also presented by various researchers including the hierarchical Fog architecture \cite{giang2015developing, hosseinpour2016approach},  OpenFog architecture \cite{openfog2016}, Fog network architecture \cite{intharawijitr2016analysis}, Fog architecture for Internet of energy \cite{baccarelli2017Fog}, Fog computing Architecture based on nervous system \cite{sun2017resource}, and IFCIoT architecture \cite{munir2017ifciot}. After reviewing the literature stated above, we define the components of Fog computing architecture, which is presented in the following subsection.

\subsection{Components of Fog computing architecture}
Fog computing architecture consists of several layers. Each layer and its components are shown in Figure \ref{Fog_Fogarc}. In this subsection, we discuss various components of the Fog computing architecture. The components are divided into several groups based on their functionality, which is defined as the layer. These functionalities will enable IoT devices to communicate with various Fog devices, servers, gateways, and the cloud. A detailed explanation of each layer is given below, where a smart transportation use case is considered in the explanation.

\Figure[t!](topskip=0pt, botskip=0pt, midskip=0pt)[width=3in]{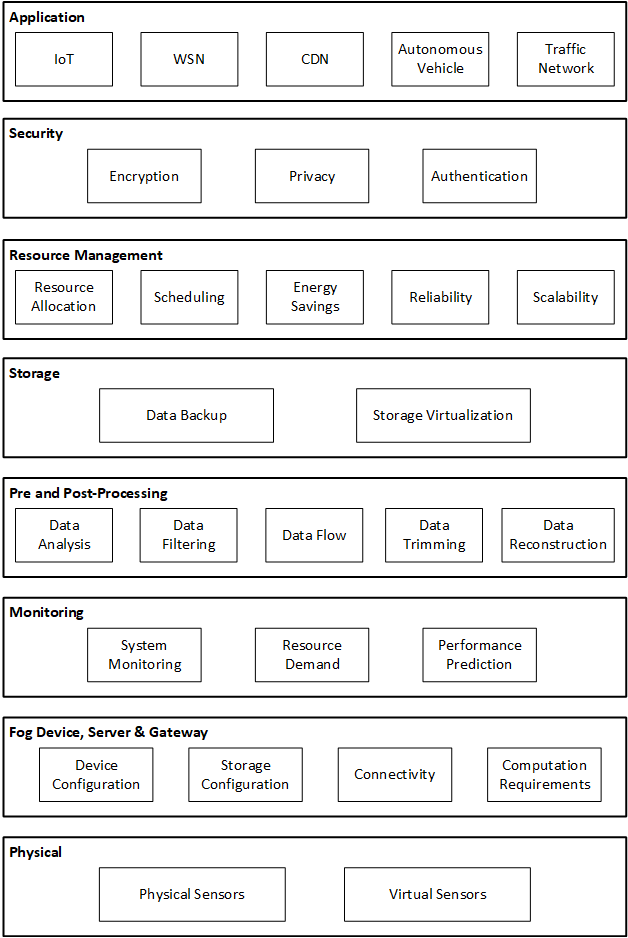} {Components of Fog computing architecture.\label{Fog_Fogarc}}


\subsubsection{Physical layer} 
The basic data source for Fog computing is the various forms of data emitted by the sensors \cite{taneja2016resource}. These data could be generated from smart devices, temperature sensors, humidity sensors, smart homes, the CCTV surveillance system, traffic monitoring system, self-driving vehicles, and so on. For instance, if we wanted to implement a smart traffic management and monitoring system, we need to get updated traffic conditions of all roads from various sensors, roadside devices, and cameras, which will help manage traffic signals. It is also necessary to predict future traffic demand by collecting data from various GPS sensors. Besides physical sensors, the role of virtual sensors is also important \cite{aazam2015Fog}, if a road accident occurred, it would not be possible to decide using just a single sensor whether the road should be blocked or traffic should keep going. The road might have one or more lanes- one lane may be affected by this occurrence while another lane could enable the traffic flow to continue, but the traffic handling capacity will be decreased due to this occurrence. In this case, a virtual sensor might help obtain an immediate decision on road conditions, traffic multiplexing, and traffic rerouting. Hence, the physical layer consists of physical and virtual sensors, where any data generation device could fall into any of these groups.

\subsubsection{Fog device, server, and gateway layer} 
The Fog device, Fog server, or Fog gateway could be a standalone device or an IoT device \cite{taneja2016resource,intharawijitr2016analysis,giang2015developing}. However, it is obvious that the Fog server should have a higher configuration than the Fog device and gateway since it manages several Fog devices. Various factors are involved so that the Fog server can run. These include its role, hardware configuration, connectivity, number of devices it can manage, and so on. Whether the Fog server is distinct or part of an IoT device is defined by its role. A group of physical and virtual sensors will be connected to a Fog device. Similarly, a group of Fog devices will be  connected to a Fog server. In this context, the Fog server should have higher processing and storage capacity compared to the Fog device. A specific cluster of Fog devices, which are connected to the same server, can communicate with each other when required. In the smart transportation use case, some application processing might depend on other Fog clusters. For example, if an application needed to find a fuel-efficient route, it might need information about other sensor clusters or Fog device clusters. To reach an appropriate decision, processing needs to be done in multiple Fog devices and servers. The Fog server and device layer are responsible for managing and maintaining information on hardware configuration, storage configuration, and connectivity of device and servers. This layer also manages the computation requirements requested by various applications. Computation requirements depend on data flow and the total number of IoT devices connected to the Fog device, as well as the total number of Fog devices connected to the Fog server. The communication between several Fog servers is maintained by this layer. For example, a Cisco IOx-supported 800 series router can be used as a Fog device and Cisco Fog data service devices can be used as the Fog server \cite{cisco_2016DS,cisco_2017FD}.

\subsubsection{Monitoring layer} 
The monitoring layer always keeps track of the system performance and resources \cite{aazam2015Fog}, services, and responses. System monitoring components help choose the appropriate resources during operation. Various applications run in smart transportation system scenarios. Therefore, it is obvious that a situation could arise when resource availability will be negative for computation or storage on a Fog device. A similar case could happen to the Fog server. To tackle these kinds of situations, the Fog device and servers will seek help from other peers. Thus, the system monitoring component will help decide such things efficiently. The resource demand component monitors current resource demand and can predict future demand for resources based on current resource usage and user activities. In this way, the system will be able to deal with any awkward situations where resource outage might occur. Performance prediction monitors can predict Fog computing performance based on system load and resource availability. This component is required to maintain appropriate QoS requirements in service level agreements. If SLA violation occurs frequently, then the cost of the system for the provider will be increased because of the penalty. Although performance prediction cannot eliminate this issue completely, it will be able to minimize overall SLA violation by predicting the performance and usage of the system. 

\subsubsection{Pre and post-processing layer} 
This layer contains multiple components, which specifically work on basic and advanced data analysis. At this level, acquired data are analyzed and filtered, and data trimming and reconstruction are also done when necessary. After processing the data, the data flow component decides whether the data needs to be stored locally or should be sent to the cloud for long-term storage \cite{giang2015developing}. The main challenge in Fog computing is to process data at the edge and minimize the volume of data that needs to be stored; this phenomenon is referred to as stream processing. In the smart transportation system use case, data will be generated from many sensors. These generated data will be analyzed and filtered in real time to get insight into the generated data. All generated data might not have any use. In some cases, it would not even be a good idea to store all generated data. As an example, if data is generated from a sensor each second, the mean value of data within a minute or within an hour may be stored depending on application requirements. Data can be trimmed in this way and a vast amount of storage space can be saved. In another case, if the difference among data values in some period of time is not that big, but might affect performance, then less numbers of reading within a minute can be taken. In such a way, it will be possible to filter a vast amount of generated data. Although the accuracy may not be 100\%, application requirements might still be fulfilled to some extent. Data reconstruction is one of the components of this layer. This module takes care of faulty and incomplete data generated by the sensors. Similarly, if one or more sensors fail during operation, this component will reconstruct the data based on the data generation pattern to prevent interruption or any other application failure.

\subsubsection{Storage layer} The storage module is responsible for storing data through storage virtualization. The data backup component ensures availability of data and mitigates the loss of data. In the storage virtualization concept, a pool of storage devices connected by a network acts as a single storage device, which is easier to manage and maintain. One of the key benefits of storage virtualization is to provide enterprise-class functionality using less-expensive storage or commodity hardware. Thus, the storage layer facilitates storage virtualization in order to minimize the complexity of the storage system. In a system, storage might fail at any point during system operation \cite{albeanu2014reliable}. Therefore, it is crucial to backup important data to mitigate any unwanted situations. The data backup module in this layer takes care of periodic or customized data backup schemes.

\subsubsection{Resource management layer} 
The components in this layer maintain the allocation of resources, and scheduling, and deal with energy saving issues. The reliability component maintains the reliability of application scheduling and resource allocation. Scalability ensures the scalability of Fog resources during peak hours where resource demand is high. The cloud deals with horizontal scalability while Fog computing aims to provide both horizontal and vertical scalability \cite{baccarelli2017Fog}. There are many distributed system resources for network, processing, and storage. This is a critical issue for distributed resources, which use application processing. Thus, resource allocation, deallocation, and reallocation will happen in which the resource allocation component manages and maintains resource allocation related issues. Another vital issue is that many applications will run in the Fog computing environment simultaneously. Hence, proper scheduling of these applications is required. The application scheduling component takes care of these applications based on various objectives. This layer also has energy saving components, which manage resources in an energy-efficient manner. Energy efficiency also positively affects the environment and helps minimize operational cost. Reliability components handle the requirement for the reliability of a system based on various reliability measures and metrics. Fog computing is a complex system that needs to take care of all IoT devices, Fog devices, and the cloud. Therefore, a device or connection might fail at any level, so reliability management is an important issue.  

\subsubsection{Security layer} 
All security-related issues such as encryption of communications and secure data storage will be maintained by the components in this layer, which also preserve the privacy of Fog users. Fog computing is intended to be deployed as a form of utility computing like cloud computing. However, in the cloud computing concept, the user connects to the cloud for services, but in the Fog computing concept the user will connect to the Fog infrastructure for the services while the Fog middleware will manage and maintain communications with the cloud. Hence, a user intending to connect to a service must be authorized by the provider. Therefore, the authentication component in the security layer processes authentication requests from users, so they can connect to the Fog computing service environment \cite{luan2015Fog}. To maintain security, it is crucial to maintain encryption between communications, so that security breaches by outsiders will not occur. The encryption component encrypts all connections from and to IoT devices and to the cloud. Fog computing components are mostly connected via a wireless connection, so security concerns are crucial. Some services in a smart city or smart house privacy are also an issue because of the involvement of user-related data in these types of systems. The Fog computing paradigm should not disclose user information without their consent. In the current age, the majority of users normally accept the security policy of the provider without reading it. Thus, special consideration of privacy should be undertaken for such services that involve user-related critical information.

\subsubsection{Application layer} 
Although the Fog was developed to serve IoT applications \cite{sarkar2016theoretical}, several other applications based on Wireless Sensor Network (WSN) and CDN also support Fog computing. Any application that has latency-aware characteristics will be able to take advantage of Fog computing. This includes any type of utility-based service that could fit within Fog computing by providing better service quality and cost-effectiveness. For example, Augmented Reality-based applications should adopt Fog computing because of its nature. It is clear that Augmented Reality will transform the modern world in the near future.  The needs of real-time processing for Augmented Reality applications can be addressed by Fog computing, which can maintain continuous improvement of Augmented Reality-related services.      

\section{Taxonomy of Fog computing}
The Fog computing taxonomy is presented in Figure \ref{Fog_tax}. This taxonomy is derived by considering existing literature and the overall viewpoint on Fog computing. The proposed taxonomy focuses on the requirements perspective for infrastructure, platform, and application.

\Figure[t!](topskip=0pt, botskip=0pt, midskip=0pt)[width=7in]{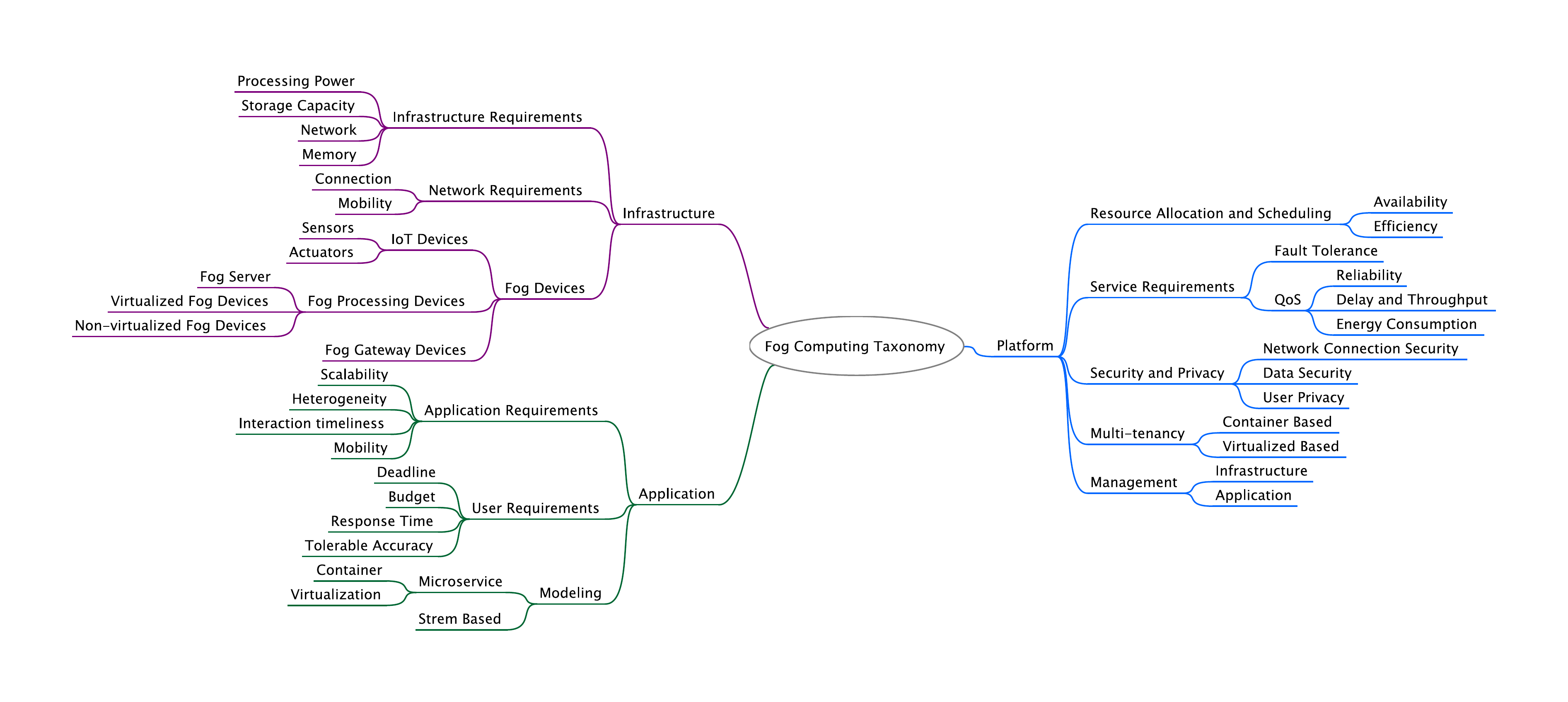} {Taxonomy of Fog computing based on the requirements of infrastructure, platform, and applications.\label{Fog_tax}}


Firstly, by considering infrastructure, we identify infrastructure and network requirements, and the types of devices in a Fog computing environment. Secondly, for platform resource allocation and scheduling, security and privacy concern, service requirements, management, and multitenancy were determined. Finally, we defined application requirements, user requirements, and application modeling taxonomy for Fog computing.
This taxonomy will help the research community and enterprises to gain better understanding and insight into the real-world deployment of Fog computing requirements, architecture, and devices. Figure \ref{Fog_tax} shows the taxonomy of Fog computing. A detailed description of each branch of the taxonomy is presented in this section.

\subsection{Infrastructure}
Fog computing infrastructure requirements depend on the network, devices, and their requirements. All Fog devices, network devices, and gateways existing in the Fog environment that participates in computation are also part of the Fog infrastructure. Infrastructure denotes the physical components of the Fog environment.

\subsubsection{Infrastructure requirements}
The many connected tiny devices are the primary elements in a Fog computing environment. These devices are located everywhere and help to connect all things around us. It is estimated that the world will see 50 billion handheld devices by 2020. Beside these devices, a huge number of sensors and actuators will also be put in place. Therefore, a proper infrastructural facility is needed to support this vast computing environment \cite{vaquero2014finding}. An example of how the number of sensors is increasing day by day is given in The Economist report titled, ``Augmented Business'', which describes the implant sensors on cattle ears that could help to monitor their activity, health, and movements. This could help increase overall productivity. The implant of sensors affixed in one cow produces about 200 MB of information in a year. In another example, with sensor technology, Rolls-Royce is able to forecast when engines will more likely fail. From such a prediction, customers can plan engine changes. Heidelberger Druckmaschinen has huge printing presses equipped with more than 1,000 sensors. These are the examples of distinct uses of sensors in specific domains. However, this phenomenon will change completely when the distinct parts are connected to generate more efficient and effective decisions. Therefore, the Fog infrastructure must have the capability to provide physical resources for computation, networking, storage, and memory to achieve efficient Fog computing services.

\subsubsection{Network requirements}
The network is one of the key bottlenecks in the Fog computing environment, where billions of connected devices generate and consume data at the edge of the network \cite{growth2013architecture}. Most of the sensing and actuating devices require low bandwidth but a higher number of devices will be connected at the same time. Therefore, existing network connection technologies like LAN, MAN, WAN, or PAN need to be investigated further and amendments will be needed to cope with the Fog computing environment to facilitate countless IoT devices. Network operators are increasingly investing in new wireless access technology research because of the number of devices per user is increasing day by day. For instance, in the cellular mobile network, base stations have a limited number of link points \cite{vaquero2014finding}. As the number of things increase, these stations will need to support increasing numbers of devices. Fog devices must act as a router for neighboring IoT devices and as a primary processing unit for IoT application in the Fog environment. Each Fog device should have a resilient connectivity to the lower and upper layer devices. Mobile ad-hoc networks could act as a basis for the Fog network because of their mobility and lower cost feature \cite{vaquero2014finding}. Hence, connection and mobility are the main requirements for the Fog network.


\subsubsection{Fog devices}
Fog computing is basically intended to support IoT-related technologies to perform processing at the edge level. Mine projects \cite{grehl2017research, turner} at the middle of the sea, airline fleets or a ship \cite{turner} can be equipped with a huge number of sensors, so it is impossible to send and store all generated data in real-time into the cloud. Some intermediate computation, processing, and services will be done by Fog computing devices. Thus, the Fog layer must have sensor management devices, Fog processing, and storage devices and Fog gateway devices. All of these devices will work collaboratively to manage and perform tasks in the Fog plane. Here, we discuss the devices that are needed for Fog computing deployment. \\


\textbf{IoT devices:} IoT devices are the devices that have sensing and actuating capability. A sensor is able to sense the environment, while an actuator acts on it when necessary. One of the most common types of sensor in IoT devices is the temperature sensor. The temperature sensor has various functions depending on different domains such as at home, in factories, and in the agriculture field. This sensor is also used to sense the temperature of soil, water, and plants in order to take proper action needed to improve service outcome. Another type of useful sensor is the pressure sensor, used especially in agriculture, smart vehicles, and aircrafts. Sensors are also used to estimate the volume of water used by the agricultural sector for cultivation and other uses. Surprisingly, a huge percentage of this water is wasted due to leaky irrigation systems and inefficient use of fresh water. Efficient use of the pressure sensor will help solve this problem. The pressure sensor is able to determine the flow of water and reduce water waste. The pressure sensor is also used in smart vehicles to determine the forces acting on it, and in aircrafts to determine altitude.  

Different groups of sensors are used for different IoT environments. For example, in healthcare, the most-used sensors are chemical, IR, pressure, and temperature sensors as well as other biosensors. On the other hand, in a smartphone, the most-used sensors are the gyroscope, GPS, and proximity sensors. 

One of the applications of the proximity sensor is to determine the presence of ear to dim or turn off the phone backlight to improve battery efficiency. This sensor is also used to monitor parking space since it can determine the presence of an object without touching it. It can also be used in a wide temperature range and is not affected by color. Its detection process also is not effected by dirt, oil, or water. There are many other sensors out there that enable IoT, which include GPS sensors, water quality sensors, level sensors, chemical/gas sensors, smoke sensors, IR sensors, humidity sensors, sound and vibration sensors, motion sensors, acceleration sensors, and machine vision sensors. There are five main types of actuators-magnetic or thermal, electrical, hydraulic, pneumatic, and mechanical. The actuator has a controlling or moving mechanism, a motor, which acts on various inputs. 

The raw application data comes from various sensors like speed sensors, cameras, temperature sensors, vehicle monitoring sensors, or GPS sensors. A typical sensor generates 10 data samples every second \cite{biswas2006assessing}. Sensors convert environmental variables such as smoke, heat, light, temperature, humidity, sound, and so on into electrical signals. These sensors are varied and can be micro-electro-mechanical systems (MEMS)-based, CMOS-based or LED sensors. Communication among sensors could be done by ZigBee, Bluetooth, Z-Wave or 6LoWPAN standards for short distance communication \cite{kocakulak2017overview}. There is a necessity for communication among sensors in some cases where one sensing output is dependent on other collective sensor outputs. These sensors will be connected to Fog devices through wireless connections. However, Fog devices collect and process data based on application requirements. 

Some example of research works based on sensors can be improved by taking advantage of Fog computing. Aziz et al. \cite{aziz2016smart} proposed a real-time health monitoring system using particular sensors in which the proposed architecture was based on GSM and GPS technologies. The system specifically monitors the body temperature and blood pressure of patients. The study used an Arduino microcontroller, dfrobot GPS/GPRS/GSM module v3.0.3, a heartbeat pulse sensor, and a lilypad temperature sensor for hardware implementation. In another study, a web-based application was developed for doctors and nurses with SMS functionality, which will be used as an emergency case. The system is able to generate GPS location, body temperature, and blood pressure. Butt et al. \cite{butt2017wearable} investigated wearable technology such as SensHand, Gloves-based system, electromyography-based and hybrid systems, leap motion, and smartwatches. The development of these kinds of technology must be integrated with the smart home system and Fog-like architecture in order to deal with emergency situations. 

Some devices such as the smartphone can be considered as both an IoT and Fog device. In the same way, if some sensors and actuators were installed in the Raspberry Pi, the device could also act as both an IoT and Fog device. 






\textbf{Fog processing devices:} Any device that has computing capability, storage, and network connectivity can act as a Fog processing device. It could be a network controller, switch, router, server, or a video surveillance camera. A Cisco 800 series router can be used as a Fog device where the IoT application can be run on the device and the device supporting Cisco IOx. To date, only Cisco 800 series routers are supporting IOx with Linux kernel with virtualization support \cite{cisco_2016IOx}. Most of these devices have a 266-400 MHz MPC8272 processor with 16 KB Cache, 64 MB random access memory and 20 MB processor board flash memory. The user can host an application on these routers. These routers have two cores-Cisco IOS runs on one core and another core is used for running IOx services.

Another type of Fog processing device is the Fog server. A Fog server can control several Fog devices in a specific domain. Cisco offers two flavors of Fog computing server deployment. One is the Cisco Fog Director, which can be deployed on any type of server with Cisco-recommended server specifications \cite{cisco_2017FD}. Another example of a Fog device manufactured by Cisco is the Fog data services, which are specifically designed for IoT [66]. However, Cisco Fog data services can only be deployed on Cisco UCS E and C Series Servers. Both will act as Fog servers; however, Cisco Fog data services are especially designed for an IoT environment. However, various organizations and bodies need to work beyond the proprietary solutions for fast technological advancement and technology adoption with a limited budget.




Fog devices and Fog servers should be deployed in such a way that any type of network management device with storage and processing capability can act as a Fog device. Similarly, the usual type of server must be able to act as a Fog server. This could be an ordinary PC since Fog is not dedicated to performing very complex processing. However, further investigation is necessary to explore the minimum system requirement for a device that can act as a Fog device or Fog server. Connectivity between Fog devices and Fog servers will be via Ethernet or wireless or a serial connection in some cases. As an example, Cisco UCS E and C Series Servers, which are generally used as Fog servers, are connected to the network via Ethernet. On the other hand, Cisco 800 series routers are connected via serial ports that support Fog computation. \\

\textbf{Gateway devices for Fog:} Many hardware boards are currently available in the market including Arduino Yun, Intel Edison, Raspberry Pi, Beaglebone Black, Arduino + Shields, Netduino, Tessel 2, and so on. These boards are currently used as IoT and gateway devices and can also be used as Fog gateway devices and as Fog devices. These boards have a built-in processor, wired and wireless adapter, and a USB port. Fog computing supports device heterogeneity, where a gateway could also be a part of the Fog computing environment. Constant et al. \cite{constant2017Fog} developed a Fog gateway using Intel Edison and Raspberry Pi. Their proposed Fog gateway integrated the data conditioning process, smart analytics, intelligent filtering, and transfer to the cloud, which needs long-term storage and temporal variability monitoring.

The IoT gateway supports various data types and communication protocols between devices and sensors. It also unifies the data format from various sensors. Current IoT gateways provide a solution for communication and do not support fully automatic configurations of newly added IoT devices \cite{kang2017internet}. Guoqiang et al. \cite{guoqiang2013design} proposed a smart IoT gateway with three key benefits. The proposed gateway has a unified external interface and pluggable architecture. It has a flexible protocol to translate various sensor data into a uniform format. The study designed a customized device with a Samsung S5PV210 mobile application processor as its gateway. However, this gateway did not have any fault tolerance or security features.

\subsection{Platform}
The platform manages applications and infrastructure in the Fog environment. It takes care of resource allocation, scheduling, fault tolerance, multi-tenancy, security, and privacy in Fog computing. Based on the taxonomy of the Fog, we discuss the requirements of the platform for Fog computing in this section.

\subsubsection{Resource allocation and scheduling}
Heterogeneous devices are the main challenges in developing proper resource allocation and scheduling in the Fog. If we wanted to use the computation power of idle devices, we need to schedule tasks on these devices efficiently. Otherwise, IoT application processing in the Fog will face complex issues, which will hinder the fulfillment of the latency awareness goal. Two of the key requirements for resource allocation and scheduling are availability and efficiency. Resources in the Fog are not dedicated, and thus availability should be ensured. On the other hand, lack of efficient resource allocation and scheduling might lead to unwanted delays in the overall processing.

\subsubsection{Service requirements}
Fog services can be defined as single or multiple user requests, where the user will constantly be updated of the outcome of the service until he or she has a subscription to that service. This means that the service outcome will not be fixed and will keep changing until the end of the service. The Fog device and Fog server perform the intermediate processing, which occurs in between user request and service output. The Fog server may communicate with the cloud for processing and information retrieval when necessary. For instance, if we considered selecting the best path based on real-time traffic in a smart transportation system, the Fog service will keep updating on the best path until the end of the journey. In this case, we need to take into account mitigation of fault, service quality, network latency, and power consumption in order to maintain the standard of the service. \\

\textbf{Fault tolerance: } Fault tolerance allows a system to keep performing even when a part of the system has failed. This failure might be software failure, hardware failure, or network failure. The solution for fault tolerance will result in a fully operational system where the system will continue its operation with a lower capability instead of shutting down totally \cite{amin2015review}. Fault tolerance is mostly investigated in the cloud \cite{poola2014robust, paul2014application, jhawar2012comprehensive, zhao2010fault, jhawar2013fault,yang2014improvement,liu2015software,mohammed2017failover,liu2016using,sampaio2017comparative,kim2017adaptive}. However, it is necessary to investigate fault tolerance in the Fog as well. Although many research works have addressed the need to explore fault tolerance issues \cite{baccarelli2017Fog,munir2017ifciot,mahmud2016Fog} in Fog computing, none have actually investigated the issue. We discuss in more detail the issue of fault tolerance in Section \ref{faultol}. \\


\textbf{Quality of service (QoS):} QoS is an important service requirement for Fog computing, which is based on reliability, network delay, throughput, and energy consumption. Besides these, resource management, power-consumption model, scheduling policy, and power failure handling are also important to ensure QoS. If some sensors fail for any reason, the accuracy of the outcome or action could be affected. Fog is intended to work with latency-sensitive systems; hence, it should maintain high reliability with a strict QoS assurance. Otherwise, the latency awareness criteria will not be fulfilled. Madsen et al. \cite{madsen2013reliability} suggested that the availability of different methodologies and algorithms work with the reliability of network connectivity and information, to ensure accuracy, which is crucial for building Fog computing-based projects.

    
\subsubsection{Security and privacy}
In this technological era, people are inevitably sharing personal information when using different applications and web services. Our personal information is no longer personal; it now belongs to many tech giants because we are using their free services on a regular basis \cite{aghasian2017scoring}. A simple example is that if anyone used an Android phone without any security settings, the built-in Android OS will automatically run GPS and map services, for which it can collect all location-related activities about the user. Therefore, information about when and which country a user has visited, where a user has dined in, which route a user uses for going to the office, home, and so on will be made available to these companies. However, these tech giants might argue that they do not disclose our data to others, as they can only see our data in our timeline only. However, a recent Facebook incident fails to convince us of the honesty of these tech giants \cite{fbinci}. 

The Fog computing paradigm is completely distributed and not intended to be centrally managed most of the time. Sensitive data might be processed in an intermediate device when the application does not have full control of the device. On the other hand, the user will not have full control over the Fog applications. Users will require more protective and innovative ways to retain their privacy and protect it from any potential and very harmful entities \cite{vaquero2014finding}. Similarly, Fog application providers also need to develop security to protect their application from unwanted data theft.

Three different types of security need to be ensured: network connection security, data security, and user privacy. Network connection security and data security are applicable from both the user and provider perspectives. Moreover, user privacy is also important because Fog processing is carried out on user data in most cases.





\subsubsection{Multi-tenancy}
Multiple tenants for the same services with an isolated runtime for each tenant are referred to as multi-tenancy service. Multi-tenancy is important for Fog because of the limited resources in a Fog environment. By enabling multi-tenancy, one instance will run in a Fog device and will serve multiple tenants (users). Multi-tenancy could be container-based or could be the usual virtualization-based. Container-based virtualization is a more lightweight and powerful virtualization solution, which the Fog can adopt, to provide the fastest processing solution. Container-based virtualization does not need to emulate the operating system to facilitate virtualization; thus, it will be easier to manage and migrate. Multi-tenancy is a requirement for the platform, and needs to be defined before deployment. Multi-tenancy may incur performance degradation and security issues \cite{syed2016pattern}]; thus, adequate and secure isolation is needed.

\subsubsection{Management}
The management of the Fog can be centralized or decentralized. Since the devices in a Fog environment belong to different domains, centralized management is not always possible. Alternatively, processing of IoT applications will be done in different Fog clusters, so management will follow a distributed nature in this case. In summary, the management of the platform in a Fog must be defined. In the case of decentralized management, similar processes must be deployed for different Fog devices to handle management issues. 

\subsection{Application}
Applications have to fulfill certain requirements to execute in a Fog environment. Here, we discuss the features required by the applications for execution.

\subsubsection{Application requirements} 

\textbf{Scalability:} The number of IoT devices are increasing very swiftly day by day all over the world, which raises a new issue of scalability. Thus, we need to deal with the scaling of devices and services in the Fog computing environment. Dependency on cloud computing has been observed for IoT application processing by many research works, where trillions of IoT devices are involved, such as that of Li et al. \cite{li2013efficient}. However, implementation of the whole application in the cloud in such an environment where IoT devices are generating a huge amount of data is neither feasible nor efficient. IoT devices are not only stationary but also mobile in most cases. Hence, maintaining frequently changing device states and availability in the cloud is not an easy task. Also, with the growing number of IoT devices, it would be more critical for IoT applications to query and select IoT devices \cite{giang2015developing}. The Fog computing system must be an autonomous system where application execution by the participating device will be done automatically including scalability. \\

\textbf{Heterogeneity:} For any IoT system, the heterogeneous device is a fundamental characteristic where device heterogeneity co-exists at any level in the Fog computing paradigm. Abstraction of device complexity is also required to some extent. Device heterogeneity does not only refer to the diversity of services and protocols, but also the assortment of horizontal and vertical levels of the Fog architecture \cite{ giang2015developing}. To address this heterogeneity, Giang et al.  \cite{giang2015developing}, classified three types of Fog devices: compute, input/output (IO), and edge nodes. Edge nodes are the sensors and the actuators, IO nodes are the resource-limited devices mostly responsible for brokering communications, and computing nodes offer computing facilities. Of the three types of nodes, IO and compute nodes are mostly dynamic and customizable or programmable as required. It is possible to implement all three nodes in a single device based on its capability and design goal. The smart gateway is an example of such implementation. In order to use the capability of various types of devices in an IoT environment, it is obvious that the application must be designed in such a way that it might be able to perform its task execution on multiple devices regardless of its capacity and location. More precisely, the application should able to use maximum available computation resources through middleware. \\

\textbf{Interaction timeliness:} The perception-action (PA) cycle is the basic function of a nervous system, which maintains circular flow between sensory organisms and its actions towards the functionality of that sensing. The PA cycle is also a characteristic of IoT applications, where the cloud and Fog infrastructure satisfies timeliness requirements and application logic for communication. Giang et al. \cite{giang2015developing} identified four interaction models for the PA cycle in a Fog environment. Examples of these models are: (i) in a local network, communication between devices, which is considered as an immediate cycle action, (ii) interaction with the cloud from a device of a local network, which is generally for time-insensitive actions, (iii) an interaction generated by the cloud to a device in a local network, which requires semi-immediate actions, and (iv) communication among IoT-related applications in the cloud. However, their work considered the role of the Fog server, which manages and maintains several Fog devices in a specific cluster. On the other hand, PA interaction can be divided into immediate, semi-immediate, and delayed action to leverage IoT application requirements more efficiently. Delayed action can be performed on the type of processing that does not have any timeliness issues and could be processed by the cloud infrastructure. \\

\textbf{Mobility:} Device mobility is a natural probability and is one of the key requirements for implementing an IoT platform \cite{giang2015developing}. From the Fog perspective, it is not only the edge devices that will be mobile but also computing and storage devices in the Fog layer. Managing mobile devices in two different planes and syncing them with each other is challenging. To ensure resource availability and successful task completion, task distribution, duplication, and migration is required. This mechanism is already considered in the cloud but there is a need to reinvestigate them by considering mobility \cite{hong2013mobile}. 

\subsubsection{User requirements}
User requirements can be changed by various constraints. First of all, a user may want to complete the submitted task within a specific time binding, which is referred to as the deadline. Secondly, the user may set some constraints for the budget. Thirdly, in the case of some users, they may not care about the budget but the response time is of utmost importance. Fourthly, some users may want  tolerable accuracy. This means that the user may not seek accurate results but rather fast results that could be provided with some reasonable errors. Aazam and Huh \cite{aazam2015Fog} suggested that pre-allocation and prediction of resources rely on user behavior and the probability of future utilization of resources. Dastjerdi et al. \cite{dastjerdi2016Fog,dastjerdi2016Fog1} stated that edge devices perform optimization by considering user behavior and network condition.

\subsubsection{Application modeling}
Two types of application modeling are possible by considering the requirements of applications in the Fog. Most IoT devices generate tuples periodically, which can be considered as a stream. These streams need to be processed in real time to get accurate results. Alternatively, the application that does the processing based on previously stored sets of data could include microservice-based applications. The advantage of microservice is that it can bind all functionality and required libraries in a single service, which can run above the microservice controller without dependency. Hence, application modeling in Fog could be stream-based or microservice-based.   

\section{Dimension of Fog computing-based applications}
Several applications require a Fog computing infrastructure to provide smooth services. These include smart transportation systems, Augmented and Virtual reality, healthcare, video streaming, smart homes, and smart cities. Requirements of platform and applications are also needed in order to provide services. In this section, we discuss some research works, which specifically address the application of Fog computing. We evaluate each work based on their contribution on the Fog infrastructure, platform, and applications as defined in our taxonomy. It is obvious that all three kinds of services are interrelated. However, each researcher only focused on one or more of these aspects. Mapping related works with our proposed taxonomy will help in finding the research gaps in Fog computing applications.  

\subsection{Smart transportation system}
Several research works have been carried out on smart transportation systems that use Fog computing. In this section we discuss a few works that have been done on the Fog-based smart transportation system and then identify key issues that need to be addressed. 

Truong et al. \cite{truong2015software} pproposed a Vehicular Ad-hoc Network (VANETs) architecture called Fog Software Defined Networking (FSDN), which combines SDN and Fog together to provide a better solution. As SDN has programmability, flexibility, global knowledge, and scalability features and Fog has location awareness and time sensitivity, the combination of these two will leverage on the key challenges in VANETs. The proposed system is able to augment communication among vehicles, infrastructure, and base stations via centralized control, besides reducing latency and optimizing the utilization of resources. However, the central SDN controller of the proposed system is where the bottleneck of the proposed system occurs. The system is focused on infrastructure and network requirements. The Fog controller is used for service implementation. The work did not focus on platform and application requirements. 

Investigation of VANETs in Fog has also been done in Giang et al. \cite{giang2016developing}, where they explored how smart transportation applications (VANETs) are developed using the Fog Computing approach. Driving vehicles in an urban area requires immediate decision on various activities such as route changing, lane change, slowing down speed, looking at obstacles, and so on. Hence, applications need to gather all related details to act on these activities. The authors discussed Fog-based smart transportation application requirements such as programming abstraction and application models. The work explored application modeling but not other application aspects nor infrastructure or platform.

\subsection{Vehicles as Fog infrastructure}
Hou et al. \cite{hou2016vehicular} proposed the idea of Vehicular Fog Computing (VFC), which will use the vehicle as an infrastructure for computation and communication. The VFC architecture utilizes vehicle computation resources by providing service to the edge devices located near them. It will aggregate abundant resources of each moving car by which service quality can be enhanced. Using quantitative analysis on different scenarios, they discovered an interesting relationship among connectivity, mobility, communication capability, and parking behavior. These four characteristics help us understand resource utilization of vehicle resources, which will help achieve better utilization of unused resources. 

\subsection{Augmented and virtual reality}
Augmented Reality applications are extremely time sensitive; a small delay can lead to serious errors in user experience. Thus, Fog computing-based solutions will have great potential in this domain \cite{dastjerdi2016Fog}. These statements are also applicable for connected Virtual Reality (VR) or VR-based games. Zao et al. \cite{zao2014augmented} proposed an augmented brain computer interaction game, which utilized the Fog and cloud infrastructure. The Fog performed real-time analysis such as signal processing that needs to classify the brain state and other analyses such as model classification updated from the cloud. However, their work only focused on the Fog infrastructure but neglected most aspects regarding platform and application. 


\subsection{Healthcare} 
The Fog computing approach also enables real-time sensor-based healthcare services. Rahmani et al. \cite{rahmani2018exploiting} proposed a Fog-assisted system architecture for the healthcare system. A smart e-health gateway is the key component of this architecture, which will process the generated data from the sensors and generate an Early Warning Score (EWS) to notify for any medical emergency. They considered many aspects of our taxonomy; but it is necessary to investigate each aspect extensively, which this study did not. Another Fog-based healthcare architecture was proposed by Mahmud et al. \cite{Mahmud2018cfog}. Their work mainly focused on network delay, power consumption, and communication optimization in Fog-based healthcare service. However, platform, application, and user requirements were not investigated.

\subsection{Smart city}
Smart city-related applications need to process sensor data on a real-time basis, where Fog computing can play a major role. Giordano et al. \cite{ giordano2016smart} proposed a Rainbow architecture, which supports various applications in a smart city. The proposed Rainbow framework evaluated three smart city applications including noise pollution mapping, urban drainage networks, and smart street. The work proposed a distributed agent-based approach in the intermediate layer in between the physical infrastructure and cloud. However, the work did not focused on application and platform aspects except for application modeling.

Table \ref{apptable} shows a summary of the above-discussed Fog-based applications mapped to our proposed taxonomy. In summary, it can be concluded that most of the works have focused on infrastructure and application modeling. There is a research gap on application- and platform-related aspects, which need to be explored further.

{\renewcommand{\arraystretch}{1.3}
\begin{table*}[htbp]
	\centering
    \scriptsize
	\caption{Evaluation of existing works on Fog applications}
	\label{apptable}
	\begin{tabular}
    {L{2cm}|L{2cm}|C{0.4cm}|C{0.4cm}|C{0.4cm}|C{0.4cm}|C{0.4cm}|C{0.4cm}|C{0.5cm}|C{0.5cm}|C{0.5cm}|C{0.5cm}|C{0.5cm}|C{0.5cm}|C{0.5cm}|C{0.5cm}} \hline \toprule
		\multirow{3}{*}{\textbf{Author \& Year}}& \multirow{3}{*}{ \textbf{Application Type}}&\multicolumn{3}{|c}{\textbf{Infrastructure}}&\multicolumn{5}{|c}{\textbf{Platform}}&\multicolumn{6}{|c}{\textbf{Application}}\\ \cline{3-16}

&&\rotatebox{90}{\multirow{2}{*}{Infrastructure requirements}}&\rotatebox{90}{\multirow{2}{*}{Network requirements}}& \rotatebox{90}{\multirow{2}{*}{Fog Devices}}&\rotatebox{90}{\multirow{2}{*}{Resource Allocation and Scheduling}}&\rotatebox{90}{\multirow{2}{*}{Service requirements}}&\rotatebox{90}{\multirow{2}{*}{Security and privacy}}&\rotatebox{90}{\multirow{2}{*}{Multi-tenancy}}&\rotatebox{90}{\multirow{2}{*}{Management}}&\multicolumn{4}{|c|}{\textbf{Application requirements}}&\rotatebox{90}{\multirow{2}{*}{User Requirements}}&\rotatebox{90}{\multirow{2}{*}{Application Modeling}} \\ \cline{11-14}

&&&&&&&&&&\rotatebox{90}{Scalability} &\rotatebox{90}{Heterogeneity}&\rotatebox{90}{Interaction timeliness}&\rotatebox{90}{Mobility}&& \\ \hline

Truong et al. \cite{truong2015software}	&	Smart Transportation System	&	\ding{51}	&	\ding{51}	&	\ding{51}	&	\ding{55}	&	\ding{55}	&	\ding{55}	&	\ding{55}	&	\ding{55}	&	\ding{55}	&	\ding{55}	&	\ding{55}	&	\ding{55}	&	\ding{55}	&	\ding{55}	\\ \hline
Giang et al. \cite{giang2016developing} 	&	Smart Transportation System	&	\ding{55}	&	\ding{55}	&	\ding{55}	&	\ding{55}	&	\ding{55}	&	\ding{55}	&	\ding{55}	&	\ding{55}	&	\ding{55}	&	\ding{55}	&	\ding{55}	&	\ding{55}	&	\ding{55}	&	\ding{51}	\\ \hline
Hou et al.\cite{hou2016vehicular}	&	Vehicles as infrastructure	&	\ding{51}	&	\ding{55}	&	\ding{55}	&	\ding{55}	&	\ding{55}	&	\ding{55}	&	\ding{55}	&	\ding{55}	&	\ding{55}	&	\ding{55}	&	\ding{55}	&	\ding{55}	&	\ding{55}	&	\ding{55}	\\ \hline
Zao et al. \cite{zao2014augmented} 	&	Augmented and virtual reality	&	\ding{51}	&	\ding{55}	&	\ding{51}	&	\ding{55}	&	\ding{55}	&	\ding{55}	&	\ding{55}	&	\ding{55}	&	\ding{55}	&	\ding{55}	&	\ding{55}	&	\ding{55}	&	\ding{55}	&	\ding{51}	\\ \hline
Rahmani et al. \cite{rahmani2018exploiting}	&	Healthcare	&	\ding{51}	&	\ding{55}	&	\ding{51}	&	\ding{55}	&	\ding{55}	&	\ding{55}	&	\ding{55}	&	\ding{51}	&	\ding{51}	&	\ding{55}	&	\ding{55}	&	\ding{51}	&	\ding{55}	&	\ding{51}	\\ \hline
Mahmud et al. \cite{Mahmud2018cfog}	&	Healthcare	&	\ding{51}	&	\ding{51}	&	\ding{51}	&	\ding{55}	&	\ding{55}	&	\ding{55}	&	\ding{55}	&	\ding{55}	&	\ding{55}	&	\ding{55}	&	\ding{55}	&	\ding{55}	&	\ding{55}	&	\ding{51}	\\ \hline
Giordano et al. \cite{ giordano2016smart}	&	Smart City	&	\ding{51}	&	\ding{55}	&	\ding{51}	&	\ding{55}	&	\ding{55}	&	\ding{55}	&	\ding{55}	&	\ding{55}	&	\ding{55}	&	\ding{55}	&	\ding{55}	&	\ding{55}	&	\ding{55}	&	\ding{51}	\\ \hline

							
        
        

        \bottomrule\end{tabular}
\end{table*}
}

\section{State-of-the-art Fog Computing}
In this section, we focus on some existing research works on Fog computing. We discuss research works from four different research areas of Fog computing. These areas are resource allocation and scheduling, failure handling, simulation tools, and microservices.

\subsection{Resource allocation and scheduling in Fog Computing}
Fog computing is fast evolving and growing rapidly due to its edge-level computation and heterogeneous nature. In this section, we present several research works, which have been done in the past couple of years. We also summarize the presented research works with a comparative discussion to address research gaps in this area. Most of the reviewed works are related to resource allocation and scheduling in the cloud and Fog environment. However, some works have only been done in the Fog environment.

\subsubsection{Resource allocation and scheduling for Fog-Cloud environment}
Alsaar et al. \cite{alsaffar2016architecture} proposed resource allocation methods for a collaborative platform composed of Fog and cloud paradigms. Their proposed algorithm is grounded on linearized decision tree rules by considering three different conditions for managing user request and for balancing workload. The conditions are VM capacity, completion time, and service size. Each condition has two branches: the VM capacity branches out to enough or not enough; the completion time consists of now or later, and the service size is divided into small or large. In some cases, this includes services in the queue, which will be represented with yes or no. They utilized 1/m/m/1, with (1)/ representing cloud broker, /(m) for many paths, /(m) for many Fog brokers, and /(1) for IoT device users. Using this method, the total overhead for big data processing in the system was reduced. In their work, the availability of cloud servers and the Fog was guaranteed and a fast response time to satisfy QoS was achieved. The SLA for users was also different, where shared and reserved resource was provided. However, availability and QoS were not studied extensively.   

Deng et al. \cite{deng2016optimal} presented a framework for workload allocation in the cloud and Fog environment to examine power consumption-delay trade-off issues. They defined the workload allocation problem into primary and sub-problems, which can be solved via related sub-systems. They employed a Hungarian algorithm and Generalized Benders Decomposition (GBD) algorithm to solve the problem. Numerical and simulation results were presented to prove that the Fog is a complement to the cloud. However, the complex nature of workload and resource was not studied in their work.

Brogi et al. \cite{brogi2017best} prototyped a tool known as `FogTorch$\prod$' which is capable of fulfilling hardware, software, and QoS requirements before deploying a composite application in the Fog infrastructure. The proposed tool manipulates Monte Carlo simulations and only considers communication link QoS. Resource consumption and QoS assurance terms were undertaken for classifying the eligibility of deployments. The proposed algorithm was based on the preprocessing phase and backtracking search phase. To find eligible deployment, the preprocessing used input from results derived by the backtracking search algorithm. However, availability and latency are more important in the Fog environment compared to resource consumption and communication links. 

In order to ensure efficient use of resources and network infrastructure in the Fog and cloud environment, Taneja and Davy \cite{taneja2017resource} proposed a Module Mapping Algorithm, which efficiently deploys IoT Application Modules in the composite Fog-Cloud Infrastructure. They employed lower-bound searches and compared function algorithms to find an eligible network node in the Fog and cloud. The Module Mapping algorithm returned a map with nodes, which are appropriate for completing the computation operation. If the application requires faster processing, the application will be deployed close to the source device. However, the work considered CPU, RAM, and bandwidth to find the best resources. In such a case, the cloud resource will always be the best resource, so it will be necessary to consider other parameters such as response time and availability of the specified resources.   

Yin et al. \cite{yin2017distributed} studied a Fog-assisted big data streaming scenario, where Fog devices are responsible for preprocessing raw data for applications hosted in the cloud using the unused resources of Fog devices. In their work, the software-defined network (SDN) controller continuously adjusted the volume of data  to be sent to the Fog device for pre-processing. The collaborative computation problem was defined as a social welfare maximization problem and a hybrid alternating direction method of multiplier (H-ADMM) algorithm was proposed to minimize computation burden via the dynamic distribution of Fog devices, cloud, and SDN using message exchanging. The formulation of social welfare maximization problem determined the size of data that will be assigned to a Fog device. During the formulation, loss of information value by preprocessing and the operation cost of the Fog and cloud were considered. The work completely depended on the cloud for post-processing, but pre- and post-processing could have been done in the Fog to support time-sensitive real-time applications. 

Aazam et al. \cite{aazam2018iot} proposed a dynamic resource estimation algorithm by integrating the historical record of cloud service customer (CSC) in a Fog environment based on the relinquish probability. The minimum relinquish probability value is 0.1 and this value will be increased based on the history of the user. However, for fair resource estimation, the relinquish probability will be 0.3 for new customers. For existing and returning customers, the characteristics of the customer are known, so the probability value can be calculated easily. In this way, resource underutilization could be minimized and the chances of profit loss will be low.

\subsubsection{Resource allocation and scheduling for a Fog environment}
A resource allocation strategy based on priced timed Petri nets (PTPNs) was proposed by Ni et al. \cite{ni2017resource} for Fog computing. The main idea of this work is that the user can choose the satisfying resources autonomously from a pre-allocated resource group. With credibility evaluation for both users and Fog infrastructure, their proposed strategy comprehensively considers the cost for time and price to complete the tasks. The user that has a high credit limit will be able to allocate highly reliable resources to complete their tasks. Due to the dynamic nature of creditability of users and resources, there will be some deviation in calculating them properly. To maintain QoS, the resources will be ordered according to their processing capacity and divided into several groups. Moreover, users with similar credibility will be assigned to several groups. 

Pooranian et al. \cite{pooranian2017novel} proposed a simple algorithm to find an optimal solution for resource allocation. They considered the problem as a bin packing penalty aware problem where servers are bins and VMs are the pack. Based on idle energy, maximum frequency, and maximum energy, each server will be palatalized and rewarded. The method will calculate how many VMs could be allocated in t time slot on a server. The VMs will be served based on their frequency and time limitations. As a consequence of penalty, a server will be punished in the form of being banned from use for a few iterations. Once the server passes the iteration freeze, it will return to the stream to perform further computation. The penalty and reward methods are applied to minimize exponentially increasing energy consumption.

Sun and Zhang \cite{sun2017resource} proposed a crowd-funding algorithm for a Fog environment, integrating idle resources in the local network. An incentive mechanism was used to encourage resource owners to participate in the computation and enthusiastically perform their tasks. Through the comprehensive reward and punishment mechanism, it is ensured that the participant will positively perform the tasks. This work is similar to the above-described literature proposed by Pooranian et al. \cite{pooranian2017novel}. However, in this case, the reward and punishment go to the participant rather than the physical server.

\subsubsection{Summary of resource allocation and scheduling in Fog}
Based on the related research on resource allocation and scheduling in the Fog, a summary is presented in Table 5. From this table, we can see that most of the researchers have focused on resource allocation in the Fog. More research works are therefore required to investigate resource sharing and workload allocation. Also, further investigation is needed to address energy-efficiency, load balancing, SLA, and QoS in the Fog. We identified two major issues in Fog computing research. Firstly, researchers tend to use a synthetic workload to validate their methods and algorithms. Secondly, most of the researchers used cloud-based simulations, which are not that convincing because the Fog is more heterogeneous and dynamic in nature. Thus, further investigation into workload generation and simulations in the Fog need to be undertaken.

{\renewcommand{\arraystretch}{1.3}
\begin{sidewaystable*}[htbp]
	\centering
    \scriptsize
    \caption{Summary of resource allocation and scheduling research in Fog}
	\label{realsctable}
	\begin{tabular}
    {L{1.8cm}|C{1.8cm}|C{1cm}|C{3.5cm}|C{1.4cm}|C{1.5cm}|C{0.4cm}|C{0.4cm}|C{0.4cm}|C{0.4cm}|C{0.4cm}|C{0.3cm}|C{2cm}|C{2cm}} \hline \toprule
		\multirow{2}{*} {Author \& year} & Proposed Algorithm	& Infrastructure & Method applied	& Evaluation Process	& Workload &\multicolumn{4}{|c}{Method proposed for} & \multicolumn{4}{|c}{Other Considerations} \\ \cline{7-14}
        &&&&&& \rotatebox{90}{Resource Allocation} &	\rotatebox{90}{Resource Provisioning}&	\rotatebox{90}{Resource Sharing}&	\rotatebox{90}{Workload allocation}& \rotatebox{90}{Energy Efficiency} &\rotatebox{90}{Load balancing} &	\rotatebox{90}{SLA Parameters} & \rotatebox{90}{QoS parameters}  \\ \cline{1-14}
        Alsaffar et al. \cite{alsaffar2016architecture}& Resource Allocation and Data Distribution	& Fog / Cloud & Utilized 1/m/m/1 model where (1) refers to cloud broker, (m) refers to many paths, (m) refers to many Fog brokers in Fog environments, and (1) refers to IoT devices users.&	Simulation (CloudSim)&	Synthetic &	Yes &	Yes &	Yes	& No	& No	& No	& Response time, target time	& Server availability, fast response time\\ \hline
        
        Deng et al. \cite{deng2016optimal} &	Generalized Benders Decomposi tion (GBD) algorithm and Hungarian Algorithm	&Fog/ Cloud	& Define optimal workload allocation among cloud and Fog which minimizes power consumption with a defined service delay. Proposed technique solved workload allocation problem in cloud and Fog.	& Simulation (MATLAB) and Numerical	& Synthetic	& No	& No	& No	& Yes	& Yes	& No	& NA	& NA \\ \hline

	Ni et al. \cite{ni2017resource} &	Dynamic resource allocation strategy	& Fog	& Ordering the resources into several groups based on their processing capabilities.	& Testbed &	Synthetic (Random)	& Yes	& No	& No	& No	& No	& No	& NA	& Price, completion time \\ \hline

	Brogi et al. \cite{brogi2017best}	& Backtracking search algorithm	& Fog/ Cloud	& Define the deployments of complex applications to Fog infrastructures, which fulfil hardware, software and service quality requirements.	& Simulation &	Yes	& No	& No	& No	& No	& No	& NA	& Latency, bandwidth \\ \hline
    
    Taneja et al. \cite{taneja2017resource}	&ModuleMapping and LowerBound Algorithm&	Fog/ Cloud	&Efficient application deployment in Fog-Cloud environment for IoT based applications by employing module mapping algorithm for efficient utilization of network infrastructure and resources. &Simulation (iFogSim)	&Synthetic	&Yes	&No	&No	&No	&Yes	&No	&NA	&Latency \\ \hline

	Yin et al. \cite{yin2017distributed}	&Distributed Resource Sharing Scheme	&Fog/ Cloud	& Grounded on an algorithm known as Hybrid Alternating Direction Method of Multipliers (H- ADMM) algorithm	&Numerical	&NA	&No	&No	&Yes	&No	&No	&No	&No	&No \\ \hline

	Pooranian et al. \cite{pooranian2017novel}	& Energy- aware algorithm for Fog data center	& Fog	&Provide reward and penalty to each server depends on their energy characteristics such as maximum frequency, maximum energy and idle energy	&Numerical and Simulation (iFogSim)	&Worldcup98 workload \cite{urgaonkar2007analytic}	&Yes	&No	&No	&No	&Yes	&Yes	&No	&Latency \\ \hline

	Sun et al. \cite{sun2017resource}	&Crowd-funding algorithm 	&Fog	&A resource-sharing model grounded on the repeated game theory in Fog computing	paradigm &Hadoop cluster	&Synthetic (Jmeter)	&No	&No	&Yes	&No	&No	&No	&SLA violation	&No \\ \hline
	
    Aazam et al. \cite{aazam2018iot}	& Dynamic resource estimation	& Fog/ Cloud	&Relinquish probabilities of the customers by considering resource utilization and the service quality experienced	&Simulation (CloudSim)	&Real IoT traces	&Yes	&No	&No	&No	&No	&No	&NA	&NA \\ \hline

        \bottomrule\end{tabular}
\end{sidewaystable*}
}

\subsection{Fault tolerance in Fog Computing}
\label{faultol}
The Fog computing paradigm is a highly distributed heterogeneous platform where the probability of device failure is very high compared to the cloud. Since the Fog is evolving, no study has yet been done on fault tolerance in Fog computing. However, fault tolerance has been mostly studied in the cloud computing paradigm.

Often, fault tolerance is measured by availability. In the cloud, faults are handled by proactive fault tolerance and reactive fault tolerance techniques at either the workflow level or task level. Reactive fault tolerance techniques are used to reduce the impact of failures on a system when the failures have actually occurred. Techniques based on this policy are job migration, checkpoint/restart, replication, rollback and recovery, task resubmission, user-defined exception handling, and workflow rescue. Proactive fault tolerance predicts the faults pro-actively and replaces the suspected components with other working components; thus, avoiding recovery from faults and errors. Proactive Fault Tolerance uses self-healing, preemptive migration, and software rejuvenation, which are the few proactive fault tolerance techniques in the cloud.

According to Sharma et al. \cite{sharma2016reliability}, the causes of failure in the cloud varies, and include software and hardware failure, service failure, overflow failure, power outage, outdated systems, network failure, cyber attacks, and human errors. It is crucial to handle faults in Fog computing for which the fault needs to be considered at every step, not only for processing but also for the transmit-and-receive process \cite{liu2017framework}. In this section, we discuss some existing research works on fault tolerance in cloud computing. We specifically focus on resource and task failure mechanisms. Then, we summarize the existing works and present a research direction for failure handling in Fog computing.

Jiang and Hsu \cite{jiang2017fault} proposed a two-level standby design for handling server failure in the cloud system. In their proposed system, cold and warm standby of the system is made available. Once any server fails, the warm standby system will replace the failed server and the failed server will be sent to the repair house. After repairing, the system will be placed in the cold standby group. The systems in the cold standby group are in a completely switched off mode. The work proposed a model to determine the necessary number of cold and warm standby systems in the cloud. However, this type of hardware failure handling is not suitable for the Fog because most of the time Fog computing devices will not be under the property of the Fog provider. Hence, task migration is the best solution for hardware failure and this should be reactive in most cases, except where the Fog device belongs to the provider. Latiff et al. \cite{latiff2017checkpointed} proposed a cloud-scheduling scheme based on a check-pointed league championship algorithm. They employed a task migration method for independent task execution failure. In their proposed method, the system state will be saved periodically by check-pointing, so the task need not start from the beginning once it fails. When the task fails, it will be assigned to an underloaded VM and the league championship algorithm will be employed to schedule the failed tasks.

Wu et al. \cite{wu2014model} proposed a fault tolerance technique using migration to the cloud. The failure handling method is proactive, which always monitors the host and continuously tries to predict the chances of failure. If the prediction becomes true, the system will look up other available resources and then migration will be performed. The proposed method will monitor CPU temperature, memory usage, and CPU fan speed, etc. To employ such a technique in Fog computing, further investigation is needed because the types of device in Fog are diverse.

A combined method of check-pointing and migration-based proactive failure handling was proposed by Egwutuoha et al. \cite{egwutuoha2013energy} for HPC and cloud. In the proposed method, the authors used a Lm-sensors open source software tool for computer health monitoring. From the monitoring data, they defined rule-based monitoring depending on temperature, fan speed, voltage, and processor utilization to predict failure. The rules are denoted as 1, 2, and 3, representing normal, warning, and critical state, respectively. They employed three different policies for migration. The first depends on the necessity lease additional node. The second removes the node, which is unhealthy based on the state. In the third, the critical state publishes to the head node. Finally, the system administrator is notified for further action. This type of approach might increase the overhead in the Fog; however, further exploration is essential. A recent study shows that proactive fault tolerance is the best solution for the cloud compared to redundant solutions \cite{sampaio2017comparative}. However, failure prediction accuracy is the key factor for these kinds of solutions. Their work considered software, hardware, and unstable behavior to predict the failure of the infrastructure. More specifically, they defined failure based on an error formula $err = (ActualTime - PredictedTime)/ActualTime \times 100\%$, which was derived from \cite{fu2010failure, sampaio2014towards}. A combination of the proactive and reactive method was applied by Gao et al. \cite{gao2014energy} to handle task failure in the cloud environment. The crash detection method and replication factor were proposed in this work to handle failures. Table \ref{tabfaulttol} shows a summary of the investigated literature on fault tolerance in the cloud. 

\begin{table*}[htbp]
    \centering
    \small
    \caption{Summary of investigated fault tolerance literature in cloud.}
    \label{tabfaulttol}
    \begin{tabular}{L{2.4cm}L{2.4cm}L{2.4cm}L{2.4cm}L{2.4cm}} \hline \toprule
        \textbf{Author \& Year} &\textbf{Types of failure} &\textbf{Failure management} & \textbf{Mechanism employed} & \textbf{Infrastructure} \\ \hline
        Jiang and Hsu \cite{jiang2017fault} 	&	Hardware failure	&	Proactive	&	Two-level standby design	&	Cloud	\\ \hline
Lati et al. \cite{latiff2017checkpointed}	&	Task failure	&	Reactive	&	Checkpointing and reallocation	&	Cloud	\\ \hline
Wu et al. \cite{wu2014model}	&	Host failure	&	Proactive	&	Migration	&	Cloud	\\ \hline
Egwutuoha et al. \cite{egwutuoha2013energy} 	&	Hardware /VM/ application failure	&	Proactive	&	Checkpointing and migration	&	Cloud/HPC	\\ \hline
Sampaio and Barbosa \cite{sampaio2017comparative} 	&	Hardware, software and unstable behavior	&	Proactive	&	Prediction	&	Cloud	\\ \hline
Gao et al. \cite{gao2014energy}	&	Task failure	&	Proactive and reactive	&	Crash detection, replication and migration	&	Cloud	\\ \hline \bottomrule
       
        \end{tabular}
\end{table*}

Because of its unstable nature of failure and heterogeneous characteristics, the hybrid failure handling method is more appropriate for the Fog computing environment.

\subsection{Simulation tools for Fog computing}
Simulation and modeling in Fog computing are still in their infancy. However, a few research works have been done on Fog computing simulation, which are focused on some specific aspect of Fog computing. Aazam and Huh  \cite{aazam2015Fog} focused on resource prediction and pricing in Fog computing. The Proposed Fog-based resource management model is able to estimate the required resources based on the probability of user behavior of future resource use. Validation and performance evaluation was done using simulation. However, they did not consider service heterogeneity, QoS, or device mobility factors. Another work proposed by Dastjerdi et al. \cite{dastjerdi2016Fog} focused on dag of the query for incident detection in a smart city use case. Both of these works used CloudSim \cite{calheiros2011cloudsim} to validate their method along with an experimental evaluation. The first toolkit for Fog simulation was developed by Gupta et al. \cite{gupta2016iFogsim}, known as iFogSim. The toolkit is used for the simulation and modeling of IoT resource management techniques in the Fog and edge computing paradigms. The most challenging problem is the design of resource management techniques, which determine analytic application distribution among edge devices, which will improve the throughput and reduce latency. The proposed simulator is capable of measuring the impact of resource management techniques in terms of network congestion, latency, cost, and energy consumption. The simulator was validated using two use cases and the authors also proposed a Fog computing environment architecture. 

Challenges in Fog computing deployment are include incorporating Fog with Emerging Technologies such as 5G Technologies \cite{gao2017Fog}, Network Function Virtualization (NFV), and Software-defined Networking (SDN). In this case, a simulator with container, SDN, and NFV support is crucial.




\begin{table*}[htbp]
    \centering
    \small
    \caption{Simulation tools used for Fog simulation and their key features.}
    \label{tabFogsim}
    \begin{tabular}{L{2.5cm}|L{2.5cm}L{6.2cm}L{2.5cm}} \hline \toprule
        \textbf{Name of the simulation tool} &\textbf{Proposed by} &\textbf{Key features} & \textbf{Usage example} \\ \hline
        
        CloudSim &  Calheiros
et al. \cite{calheiros2011cloudsim} & A broad simulation toolkit that enables simulation and modeling of application provisioning in the cloud computing environments. The CloudSim toolkit supports system modeling of cloud system components such as virtual machines (VMs), data centers and resource allocation and provisioning policies as well as support system behavior modeling. & Aazam and
Huh \cite{aazam2015Fog}; Dastjerdi
et al. \cite{dastjerdi2016Fog} \\ \hline
        
        iFogSim & Gupta et al. \cite{gupta2016iFogsim} & Modelled IoT and Fog environments and measure the impact of resource management techniques in terms of network congestion, latency, cost and energy consumption. Developed on top of CloudSim & Baccarelli
et al. \cite{baccarelli2017Fog}; Bittencourt
et al. \cite{bittencourt2017mobility}; Markakis
et al. \cite{markakis2017exegesis}; Taneja and
Davy \cite{taneja2017resource}  \\ \hline \bottomrule
        
        \end{tabular}
\end{table*}

Table \ref{tabFogsim} presents the key features of these two simulators that are mostly used by various researchers for Fog computing simulation. These two simulation tools did not focus on network parameters such as bandwidth distribution of the link and round-trip delay of the various media. These two parameters heavily affect the simulation results where minimization of latency is the key goal in a Fog computing environment. Secondly, both tools did not consider container-based virtualization. In a Fog computing environment, there are many devices that will participate in computation, where hypervisor-based virtualization is nearly impossible to implement due to the lower memory and processing power of these devices.

\subsection{Fog-based micro services}
A microservice is an independent process and Software-Oriented Architecture (SOA) that interacts by message passing. The SOA of microservice does not hinder or favor any specific programming model. It provides design and implementation guidelines for distributed applications to partition each component independently. Each of the components addresses a specific functionality. The functionality of the components can be accessed by message passing and is possible to implement in any mainstream programming language internally. In this way, this principle helps developers and project managers to develop each module independently and test it with a few related functions. Some microservices, also known as high-level microservices, are mainly responsible for coordination with other microservices \cite{dragoni2017microservices}. The organizational approach of microservices accelerates the development cycle, nourishes ownership, encourages innovation, improves scalability, and enhances the maintainability of software applications. Using this approach, software becomes a small independent service and interacts over unambiguous APIs. These services are preserved via self-contained small teams \cite{jung2016microservices}. 
   
The agility and independent distributed nature of microservice deployment makes it a good solution for Fog-based IoT application development. Independent processes and interaction via message passing features has made microservices more convenient for IoT applications. In the Fog, there is a limitation of resources, so developing microservices in the Fog will minimize the growing complexity of the big system by dividing it into a set of small independent services. Microservice is taking modularity to a subsequent level by incorporating high cohesion and loose coupling of distributed systems.

\subsubsection{Current research aspects of microservice}
Recently, microservice-based applications have started gaining popularity \cite{ dragoni2017microservices}. Fog-based microservices have not been investigated extensively; hence, it is an open research area. However, some research works have been done in this emerging research area, with most of the efforts being related to IoT. Butzin et al. \cite{butzin2016microservices} investigated the use of microservices in IoT and claimed that the architectural goal of IoT and microservices are similar. However, they actually have different features in terms of various aspects. First of all, microservice has a self-containment feature where all dependencies and libraries are packed with the application in a single image. On the other hand, for IoT, all libraries are not generally wrapped with the application. However, both use similar types of virtualization and web protocols. Microservice also has a continuous integration and delivery feature while in IoT these are not available or only partly exist. 

Vresk and \v{C}avrak \cite{vresk2016architecture} proposed a microservice-based middleware for IoT to support device heterogeneity, various communication protocols, and services. They presented a data model and address model for microservice-based IoT. Brito et al. \cite{de2017service} proposed a service orchestration architecture for Fog using microservices. The authors defined the resource manager as a microservice. Khazaei et al.  \cite{khazaei2017end} proposed a generic programmable self-managing microservice-based platform for IoT. In the platform, microservices will exist in all layers in a cascading manner and an autonomic management system will scale the microservice. A similar type of IoT framework was proposed by Sun et al. \cite{sun2017open}. In their architecture of nine components, all are microservices except for the core service. The work proved that microservices are far better than the monolithic approach in terms of scalability, flexibility, and platform independence. However, still, microservice-based IoT architecture suffers from various issues such as faults in the network, network delay, message serialization, cooperative transaction processing, and other distributed computing scenarios. Li et al. \cite{li2017cooperative} proposed a cooperative-based model specifically for smart sensing devices; it is possible to enhance the performance of such a service by undertaking a micro-service based concept. Krivic et al. \cite{ krivic2017microservices} proposed a management solution for Machine-to-Machine (M2M) device communication in an IoT system using collaborative microservice. They argued that microservices could act as the agent in an agent-based system since each microservice is responsible for performing a specific task and acts collaboratively to achieve the system goal. 

Container-based virtualization is the best solution for deploying microservices since the container supports OS virtualization and packs all dependencies in a single image. A container is able to manage physical hardware resource needed by an application with its OS kernel utilities \cite{ xavier2013performance}.

\subsubsection{Microservices and IoT applications}
Many research works have suggested Fog-based processing for IoT applications in smart transportation systems \cite{truong2015software, hou2016vehicular, giang2016developing}, Augmented and Virtual Reality \cite{ zao2014augmented, kosta2012thinkair, bastug2017toward}, and healthcare \cite{ cao2015fast, stantchev2015smart,Mahmud2018cloudFog}. Fog computing is also suitable for video streaming, smart homes, smart cities, and CDN. The common characteristics of these applications are time-sensitiveness, which make Fog computing a promising emerging computing paradigm. The main drawback of the Fog, however, is resource limitation and failure. Thus, using microservices for Fog-based IoT applications will minimize these drawbacks. Microservices are standalone, lightweight, and easily deliverable. To mitigate resource limitation, the microservice-based container is the best solution so far. In the same way, it will also minimize the cost of failure by deploying the application immediately. Many open research issues can be addressed by implementing Fog-based micro services; these include service management, scheduling, monitoring, fault tolerance, security, and privacy.



\subsection{Fog based mobile computing}
The number of smartphone and mobile device users in urban areas as well as in rural areas is increasing day by day. As a result, mobile users are now requesting high-volume content collaboratively. Providing service for all requested contents in an area where mobile users are densely populated is a really challenging task for service providers \cite{khan2017offloading}. The high number of concurrent content requests will only make the situation worse. One of the best solutions to cope with this problem is to offload content near the users, so that the users could get better service. This content offloading process can be supported by mobile Fog computing. In mobile Fog computing, content will be offloaded to the Fog device, which is located closer to the users. However, content management in Fog nodes is a current research issue. Depending on the demand of the contents, offloading should be distributed on the Fog nodes. Constant monitoring and efficient cache management is crucial to deal with resource-limited Fog nodes. A few research works have been done on mobile Fog computing. This section chronologically discusses the research works that have been done in this particular research area in the past couple of years. 

Hong et al. \cite{hong2013mobile} proposed a high-level programming model that supports large-scale geospatially distributed time-sensitive applications. The proposed Mobile Fog programming model has two design goals. The initial goal is to provide a simplified application development for an enormous number of heterogenous devices, which are distributed in a wide area. The next goal is the dynamic scaling of resources based on resource demand. They developed an API for their programming model and evaluated it using two application models: vehicle tracking using a camera and traffic monitoring using a Mobility-driven distributed Complex Event Processing (MCEP) system. However, they did not focus on process placement or process migration.

Shi et al. \cite{shi2015combining} proposed a P2P inspired communication model between the mobile device cloud and mobile nodes to share resources and computation task among mobile devices. In their work, they utilized a Constrained Application Protocol (CoAP) for implementing microservices. This work introduced the M2M approach in Fog computing while the classical Fog is actually hierarchy based. Content offloading in the mobile Fog was investigated by Khan et al. \cite{khan2017offloading}. They defined mobile Fog as co-located self-organizing mobile nodes, which offer distributed resources at the edge. The aim of this work was to collaborate nodes for content caching, which will maximize the availability of the content and minimize operational cost. The proposed coalition game helps find the best co-located candidates near the users for sharing storage and self-organizing.

Wang et al. \cite{wang2017data} proposed a three-layer hierarchy framework using a Fog structure to bridge the communication between WSNs and the cloud. They designed a routing algorithm for bridging communication by considering the number of hops and energy consumption. They defined the Fog node as a sink, which will transfer data from sensor to the cloud. The proposed framework consists of routing layer, Fog layer, and sink layer. In the Fog layer, a sink acts as the Fog nodes as well to minimize transmission delay. However, there is a lack of security and privacy concern being addressed in mobile Fog computing. Roman et al. \cite{roman2018mobile} addressed security and privacy for all edge-level computing. This included a usual thread in a network system mobile Fog that requires extra measures of authentication, trust, access control, protocol, and network security.

\section{Open issues and future research directions}
\subsection{Infrastructure-related issues}
The Fog is an evolving technology, expanding in such a way that it needs to reach market adoption to support all kinds of time-sensitive applications. The Fog has become an enactment of research efforts by various academies and industries. One of the key initiatives is the Open Fog Consortium (OpenFog), which was founded by the ARM, Cisco, Dell, Intel, Microsoft and Princeton University in November 2015 \cite{chiang2016Fog}. Foxconn, General Electric, Hitachi, Sakura Internet, ShanghaiTech University, and ZTE are the contributing members of this nonprofit consortium. They are accelerating digital innovation with the blending of 5G wireless technology, IoT, and embedded AI by providing open interoperable architecture. However, many open challenges exist for this sprout-level computing paradigm. In this section, we discuss the research challenges and address the future directions for Fog computing research. Figure \ref{Fog_resis} shows some important research issues in Fog computing. 

\Figure[t!](topskip=0pt, botskip=0pt, midskip=0pt)[width=3.3in]{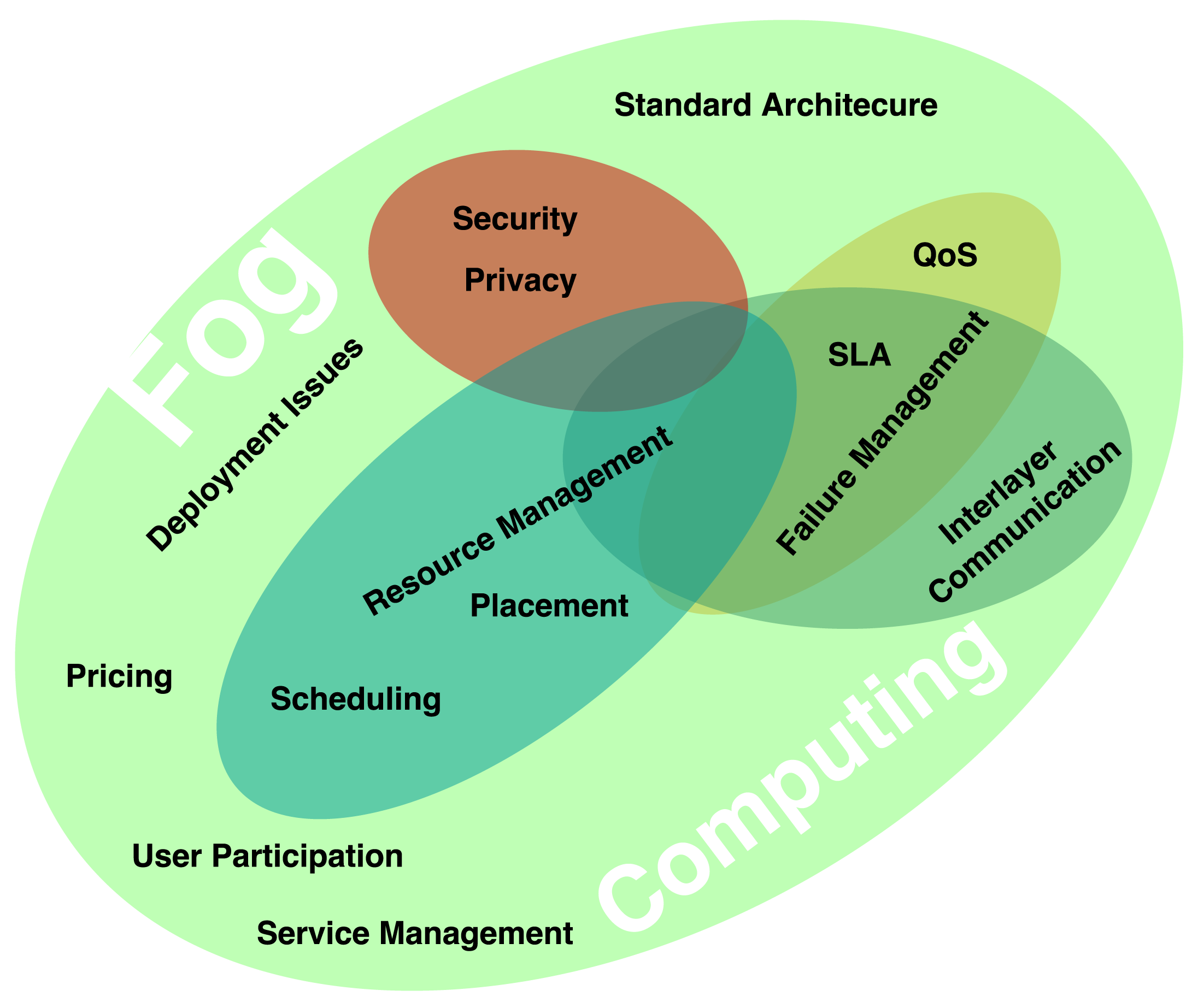} {Fog computing research issues.\label{Fog_resis}}


\subsubsection{Deployment issues}
From the deployment viewpoint, OpenFog is defined as an N-tier environment. However, the excessive increase in number of levels in the Fog layer might cause latency problems in the newly emerging Fog computing paradigm. Therefore, the number of tiers based on the use case must be determined. Deployment decisions will be undertaken based on requirements such as type and amount of task that will be done by each tier, total number of sensors, Fog device capability, in between the latency and reliability of Fog devices. Still, it is necessary to investigate how these requirements will be fulfilled. Application and resource scaling is also an important issue during deployment. Based on the requirement of the application and resource, scaling and shrinking without interrupting current services could be undertaken. In this regard, placement might also affect Fog deployment.

\subsubsection{Standard architecture for Fog computing}
Up until now, there has been no defined standard architecture for Fog computing. The OpenFog consortium released two versions of the Fog architecture in February 2016 \cite{openfog2016} and February 2017 \cite{openfog2017openfog}. Their first draft was an initial overview of the Fog architecture. In their second draft, the Fog architecture was discussed in more detail. In their proposed architecture, they considered many key aspects of Fog architecture including performance, manageability, security, data analytics, and control. However, further research needs to be undertaken to explore and gain deeper insight into each layer with proper validation. 

\subsubsection{Interoperability and federation of Fog}
Because of the Fog, users are able to process their request near them, which will minimize latency. However, what will happen if an increasing number of multiple latency-aware applications requests are sent in one shot to the Fog device and the Fog device is unable to handle that many requests? Will it be passed to the cloud for processing? If it is passed to the cloud, then latency requirements will not be satisfied. Thus, interoperability and federation among Fog clusters and Fog servers are necessary. Hence, if a Fog device is fully utilized, it will send requests to peer Fog devices or Fog servers for processing instead of sending them to the cloud.   

\subsection{Platform-related issues}
\subsubsection{Resource management}
Resources are most dynamic and heterogeneous in a Fog environment because of the diversity of devices and their available resources. All devices known as Fog devices are responsible for performing the computation of their own application. For example, a computer that relies on office staff to perform some ordinary email and documentation works might be a part of the Fog and might also act as a Fog device. In such a case, the amount of resources available for Fog computation is dynamic but predictable via the analysis of the long-term activity of its resources. This prediction is necessary because once the Fog task execution starts, and over a period of time, the status of the resources might change due to the request by the application for which the device is responsible for. If we compared this to the cloud, it is possible to know how much resources are currently available and whether or not they are exclusively used for cloud-based application requests. However, the Fog aims to use idle resources available on any Fog device with Fog computation always taking second priority. Hence, resource allocation and scheduling in the Fog is more challenging than traditional resource allocation and scheduling in the cloud. 

\subsubsection{Failure management}
Fog device failure probability is always high because the devices are distributed and the management of Fog devices is not central. Hence, the devices could fail for many reasons; this could be due to hardware failure, software failure, or because of user activity. Besides these problems, some other reasons include connectivity, mobility, and power source, which also play a big role. Most of the devices in a Fog environment might be connected via wireless connections; it is obvious that wireless connections are not always reliable. The majority of devices that are connected via wireless are mobile, so these devices could change location to different clusters frequently. One other characteristic of these devices is that they are battery powered and might fail anytime. Hence, dealing with the complex nature of failure is very difficult. Also, it is necessary to ensure SLA by defining QoS parameters. So, the question is: What are the SLAs and how should they be defined? Also: What QoS parameters must be considered, so that the consumer and providers can retain a win-win situation? 

\subsubsection{Communication between different layers}
The Fog should ensure uninterrupted connection with the devices to ensure application requirements of time-sensitive applications are met. If the application were to control an autonomous car or drone and if it were responsible for emergency surveillance, then failure in connectivity might cause serious harm. Even if connectivity to the cloud fails, the Fog still needs to ensure continuous connectivity. Thus, cross-layer connectivity among IoT devices, Fog, and cloud are of the utmost importance. The connection type and protocols used by IoT devices and Fog devices might be different. Therefore, how these issues will be handled is an important research issue.

\subsubsection{User participation management}

Efficient Fog service management depends on the participation of users in Fog computation. However, how can user participation be managed? How do we deploy minimum resources in the case where no one wants to participate? We need to address these problems clearly with a feasible solution. One of the methods to increase user participation is through incentive and reward-based policies. With such policies, any user that participates in Fog computation will benefit. Even a user, who participates to complete his own request, will be rewarded by getting some discount based on his participation. However, this area needs to be addressed because the overall success of Fog computation depends on the participation of Fog devices, which are owned by various people and organizations. 




\subsubsection{Security and privacy}
Fog devices are managed by different operators based on their location and ownership. Nobody would want to contribute to Fog computation if device control were compromised. Thus, how the security of a participant device will be maintained if the device were to take part in Fog computation is a big question. Another key security issue for this scenario is participant user data security. A participant device might have critical information. How will safety be guaranteed in such a case? On the other hand, critical data might be processed in a device, which is owned by a black hat hacker. How will safety and privacy be ensured then? Security issues also exist during cross-layer communication. Similar to the distributed nature of the Fog, security management should also be distributed, which will not be dependent on any central component.

\subsection{Application-related issues}
\subsubsection{Application service management}
Billions of IoT devices will be handled by the Fog paradigm, which will handle time-sensitive and time-insensitive applications. The degree of service, availability, and quality is most diverse in the Fog. Hence, service management is a typical issue for the entire Fog realm. Services should be microservice-based, so that agility and management issues can be handled properly. Further research is necessary to explore the possibility of Fog-based solutions.  
\subsubsection{Application modeling}
Modeling Fog applications is complex because the application should collect data from different IoT devices, which use different protocols and sets of codes. Thus, it is challenging to model generic applications, which can be deployed with minimal effort. To solve this issue, a standard form of communication protocol is necessary, so that the modeled application can communicate and work with different types of IoT devices.

\section{Conclusion}
The Fog computing paradigm is currently in its infancy, so an extensive investigation is required for this emerging technology. In this survey, we presented and discussed the overview, architecture, state-of-the-art and other similar technologies in Fog computing. Based on the literature, we derived a taxonomy for Fog computing by analyzing the requirement of Fog infrastructure, platform, and applications. We also covered resource allocation and scheduling, fault tolerance, simulation tools, and microservices in Fog computing. Finally, we presented some challenging and open research issues. We strongly believe that this comprehensive survey will bring to light IoT application execution for a Fog computing environment as well as point towards the direction for current and future research in this rapidly growing research area. In this way, this computing paradigm, which is still immature, will be propelled towards achieving market adoption in the near future.

\bibliographystyle{IEEEtran}
\bibliography{access}

\begin{IEEEbiography}
[{\includegraphics[width=1in,height=1.25in,clip,keepaspectratio]{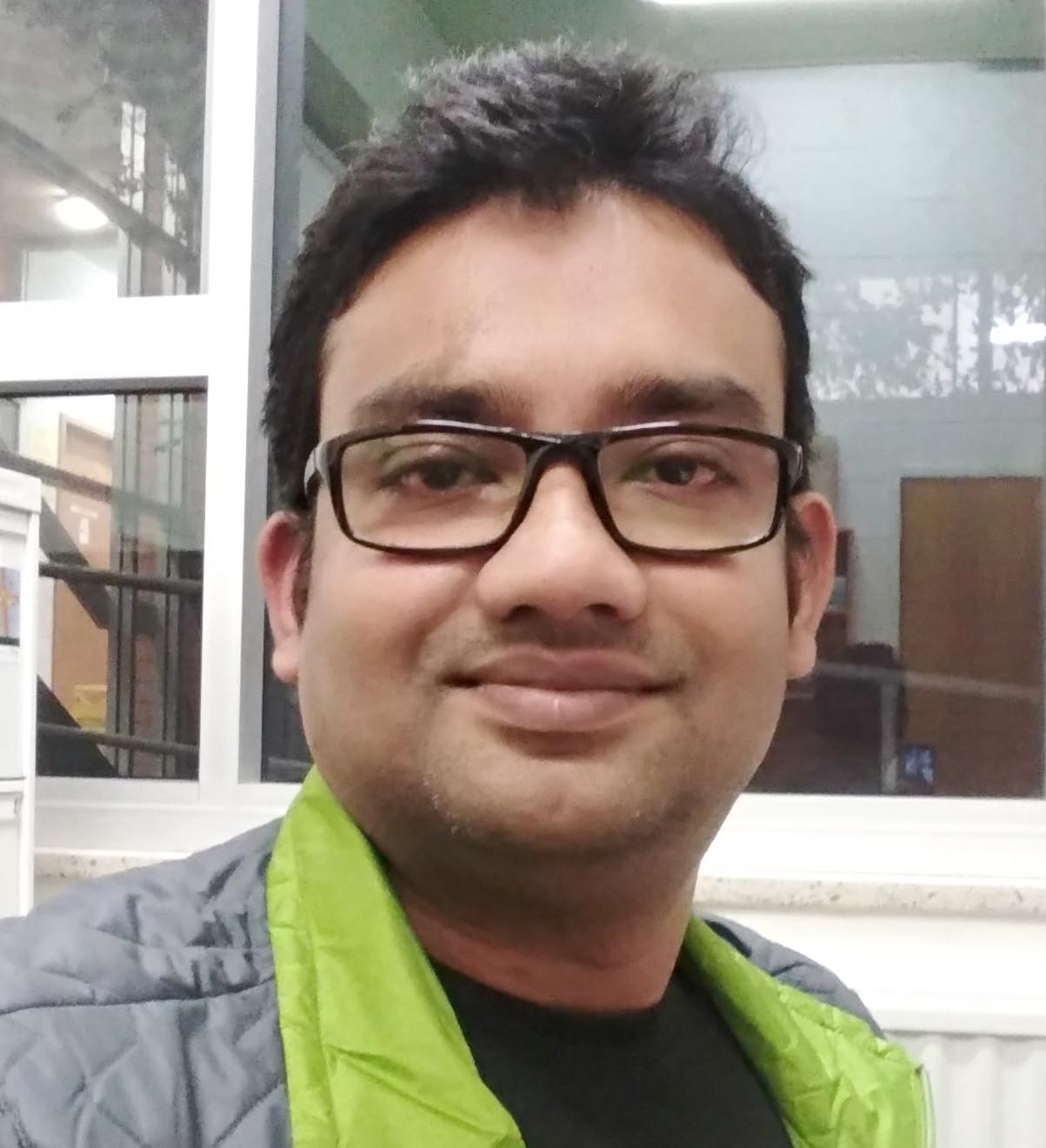}}]{Ranesh Kumar Naha} received his Master of Science (M.Sc.) degree from Department of Communication Technology and Network, Faculty of Computer Science and Information Technology, Universiti Putra Malaysia, in 2015. He received B.Sc. degree in Computer Science and Engineering from State University of Bangladesh in 2008. He is currently pursuing his Ph.D. studies on reliable resource allocation and scheduling in Fog computing environment with the University of Tasmania. He has been awarded Tasmania Graduate Research Scholarship (TGRS) for supporting his studies. His research interests  include wired and wireless network, parallel and distributed computing, Cloud computing, Internet of Things (IoT), and Fog computing. During his master study he has been awarded Commonwealth Scholarship provided by Ministry of Higher Education, Malaysia. He served as Lecturer until 2011 in Daffodil Institute of IT, Bangladesh.
\end{IEEEbiography}

\begin{IEEEbiography}[{\includegraphics[width=1in,height=1.25in,clip,keepaspectratio]{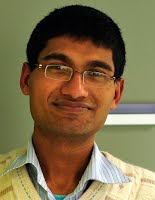}}]{Dr. Saurabh Garg} is currently a Lecturer with
the University of Tasmania, Australia. He is one
of the few Ph.D. students who completed in less
than three years from the University of Melbourne.
He has authored over 40 papers in highly cited
journals and conferences. During his Ph.D., he has
been received various special scholarships for his
Ph.D. candidature. His research interests include
resource management, scheduling, utility and grid
computing, Cloud computing, green computing,
wireless networks, and ad hoc networks
\end{IEEEbiography}

\begin{IEEEbiography}[{\includegraphics[width=1in,height=1.25in,clip,keepaspectratio]{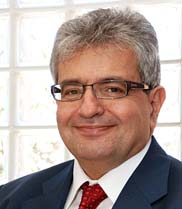}}]{Prof. Dimitrios Georgakopoulos} is a Professor in Computer Science and Director of the Key Lab for IoT at Swinburne University
of Technology, Melbourne, Australia. Before that
was Research Director at CSIRO's ICT Centre and Executive
Director of the Information Engineering Laboratory, which
was the largest Computer Science program in Australia.
Before CSIRO, he held research and management positions
in several industrial laboratories in the US, including
Telcordia Technologies (where he helped found two of
Telcordia's Research Centers in Austin, Texas, and Poznan,
Poland); Microelectronics and Computer Corporation
(MCC) in Austin, Texas; GTE (currently Verizon) Laboratories in Boston, Massachusetts;
and Bell Communications Research in Piscataway, New Jersey. He was
also a full Professor at RMIT University, and he is currently an Adjunct Prof. at the
Australian National University and a CSIRO Adjunct Fellow. Prof. Georgakopoulos
has produced 170+ journal and conference publications in the areas of IoT, process
management, and data management, and has 10,500+ lifetime citations.
\end{IEEEbiography}

\begin{IEEEbiography}[{\includegraphics[width=1in,height=1.25in,clip,keepaspectratio]{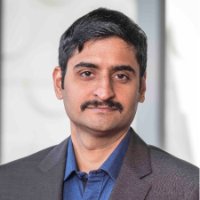}}]{Dr. Prem Prakash Jayaraman} is currently a Research Fellow
at Swinburne University of Technology, Melbourne. His
research areas of interest include, Internet of Things, cloud
computing, mobile computing, sensor network middleware
and semantic internet of things. He has authored/coauthored
more than 50 research papers in international
Journals and conferences such as IEEE Trans. on Cloud
Computing, IEEE Selected areas in Communication, Journal
of Computational Science, IEEE Transactions on Emerging
Topics in Computing, Future Generation Computing
Systems, Springer Computing, ACM Ubiquity Magazine,
IEEE Magazine. He is one of the key contributors of the Open Source Internet of
Things project OpenIoT that has won the prestigious Black Duck Rookie of the Year
Award in 2013. He has been the recipient of several awards including hackathon
challenges at the 4th International Conference on IoT (2014) at MIT media lab,
Cambridge, MA and IoT Week 2014 in London and best paper award at HICSS
2016/2017 and IEA/AIE-2010. Previously he was a Postdoctoral Research Fellow at
CSIRO Digital Productivity Flagship, Australia from 2012 to 2015.
\end{IEEEbiography}

\begin{IEEEbiography}[{\includegraphics[width=1in,height=1.25in,clip,keepaspectratio]{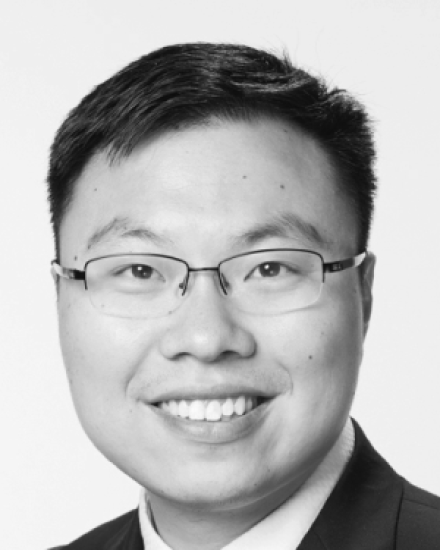}}]{Dr. Longxiang Gao} received his PhD in Computer Science from Deakin University, Australia. He is currently a Lecturer at School of Information Technology, Deakin University. Before joined Deakin University, he was a post-doctoral research fellow at IBM Research \& Development Australia. His research interests include data processing, mobile social networks, Fog computing and network security.

Dr. Gao has over 30 publications, including patent, monograph, book chapter, journal and conference papers. Some of his publications have been published in the top venue, such as IEEE TMC, IEEE IoT, IEEE TDSC and IEEE TVT. He received 2012 Chinese Government Award for Outstanding Students Abroad (Ranked No.1 in Victoria and Tasmania consular districts). Dr. Gao is a Senior Member of IEEE and active in IEEE Communication Society. He has served as the TPC co-chair, publicity co-chair, organization chair and TPC member for many international conferences.

\end{IEEEbiography}
\begin{IEEEbiography}
[{\includegraphics[width=1in,height=1.25in,clip,keepaspectratio]{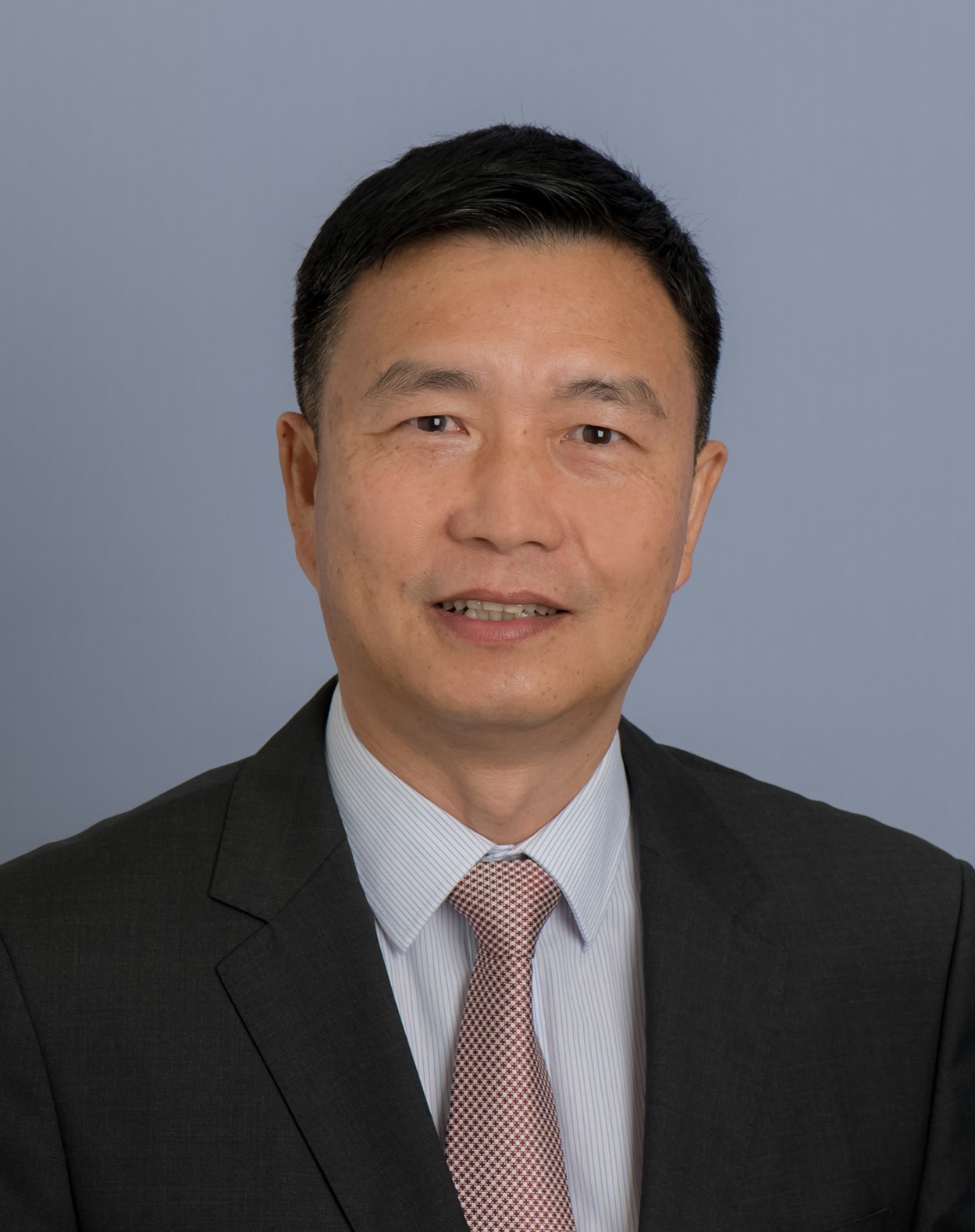}}]{Dr. Yong Xiang} 
received the Ph.D. degree in Electrical and Electronic Engineering from The University of Melbourne, Australia. He is a Professor and the Director of the Artificial Intelligence and Image Processing Research Cluster, School of Information Technology, Deakin University, Australia. His research interests include information security and privacy, signal and image processing, data analytics and machine intelligence, and Internet of Things. He has published 2 monographs, over 100 refereed journal articles, and numerous conference papers in these areas. He is an Associate Editor of IEEE Signal Processing Letters and IEEE Access. He has served as Program Chair, TPC Chair, Symposium Chair, and Session Chair for a number of international conferences
\end{IEEEbiography}
\begin{IEEEbiography}[{\includegraphics[width=1in,height=1.25in,clip,keepaspectratio]{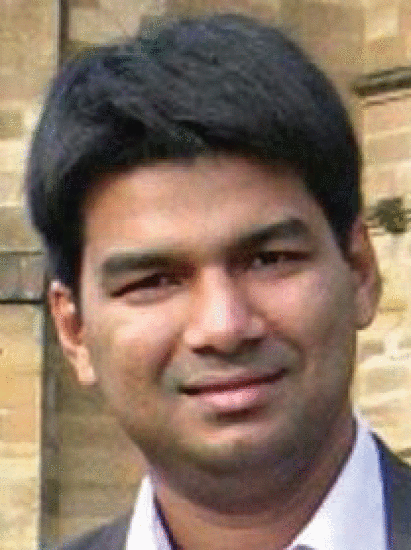}}]{Dr. Rajiv Ranjan} is a reader at Newcastle University, UK. Prior to this position, he was a Research Scientist and a Julius Fellow in CSIRO Computational Informatics Division (formerly known as CSIRO ICT Centre). His expertise is in datacenter cloud computing, application provisioning, and performance optimization. He has a PhD (2009) in Engineering from the University of Melbourne. He has published 62 scientific, peerreviewed papers (7 books, 25 journals, 25 conferences, and 5 book chapters). His hindex is 20, with a lifetime citation count of 1660+ (Google Scholar). His papers have also received 140+ ISI citations. 70\% of his journal papers and 60\% of
conference papers have been A*/A ranked ERA publication. Dr. Ranjan
has been invited to serve as the Guest Editor for leading distributed systems
journals including IEEE Transactions on Cloud Computing, Future
Generation Computing Systems, and Software Practice and Experience.
One of his papers was in 2011's top computer science journal, IEEE
Communication Surveys and Tutorials

\end{IEEEbiography}

\EOD

\end{document}